\documentclass[preprint,prd,aps,tighten,nofootinbib,amssymb]{revtex4}
\usepackage{graphicx}
\usepackage{epsfig,amssymb,amsmath,color}

\newcommand{\beq}{\begin{equation}}
\newcommand{\eeq}{\end{equation}}
\newcommand{\bea}{\begin{eqnarray}}
\newcommand{\eea}{\end{eqnarray}}
\newcommand{\bear}{\begin{array}}
\newcommand {\eear}{\end{array}}
\newcommand{\bef}{\begin{figure}}
\newcommand {\eef}{\end{figure}}
\newcommand{\bec}{\begin{center}}
\newcommand {\eec}{\end{center}}

\begin{document}
\draft
\tighten
\preprint{TU-942}
\title{\large \bf
Higgs phenomenology in the Peccei-Quinn invariant NMSSM
}
\author{
    Kiwoon Choi\,$^{a}$\footnote{email: kchoi@kaist.ac.kr},
    Sang Hui Im\,$^{b}$\footnote{email: shim@phya.snu.ac.kr},
    Kwang Sik Jeong\,$^c$\footnote{email: ksjeong@tuhep.phys.tohoku.ac.jp},
    Min-Seok Seo\,$^d$\footnote{email: minseokseo@apctp.org }}
\affiliation{
 $^a$ Department of Physics, KAIST, Daejeon 305-701, Korea \\
 $^b$ Department of Physics and Astronomy and Center for Theoretical Physics,
      Seoul National University, Seoul 151-747, Korea \\
 $^c$ Department of Physics, Tohoku University, Sendai 980-8578, Japan\\
 $^d$ Asia Pacific Center for Theoretical Physics, Pohang, Gyeongbuk 790-784, Korea
    }

\vspace{2cm}

\begin{abstract}

We study the Higgs phenomenology in the Peccei-Quinn invariant NMSSM (PQ-NMSSM) where the low energy mass parameters of the singlet superfield are induced by a spontaneous breakdown
of the Peccei-Quinn symmetry.
In the generic NMSSM, scalar mixing among CP-even Higgs bosons is constrained by
the observed properties of the SM-like Higgs boson, as well as by the LEP bound on
the chargino mass and the perturbativity bound on the singlet Yukawa coupling.
In the minimal PQ-NMSSM, scalar mixing is further constrained due to the presence of
a light singlino-like neutralino.
It is noticed that the $2\sigma$ excess of the LEP $Zb\bar b$ events at $m_{b\bar b}\simeq 98$
GeV can be explained by a singlet-like 98 GeV Higgs boson in the minimal PQ-NMSSM with
low $\tan\beta$, stops around or below 1~TeV, and light doublet-higgsinos
around the weak scale.

\end{abstract}

\maketitle

\section{Introduction}

There are many reasons to anticipate new physics beyond the standard model (SM), including the
naturalness problems such as the hierarchy problem and the strong CP problem, and a variety of
cosmological observations such as the existence of dark matter, the matter-antimatter asymmetry, and
the evidences for inflation in the early Universe.
Among the known scenarios of new physics, a particularly compelling possibility is a supersymmetric
extension of the SM \cite{nilles} incorporating also the axion solution to the strong CP problem
through a spontaneously broken Peccei-Quinn (PQ) symmetry \cite{PQ-mechanism,PQ-review}.
While solving the two major naturalness problems of the SM, such an extension of the SM provides
an attractive candidate for dark matter, either the lightest supersymmetric particle or
the axion, or both.
It also offers an interesting possibility that the PQ scale is generated by an interplay between
supersymmetry (SUSY) breaking effect and Planck-scale suppressed effect, yielding an intermediate
PQ scale $v_{\rm PQ}\sim \sqrt{m_{\rm soft}M_{Pl}}$ in a natural manner \cite{axion-scale,axion-thermal},
where $m_{\rm soft}$ is a soft SUSY breaking mass presumed to be of the order of the weak scale.
In such a scenario, the PQ phase transition takes place in the early Universe at a temperature
$T\sim m_{\rm soft}$.
This results in a late thermal inflation over the period $m_{\rm soft}< T< v_{\rm PQ}$, with which dangerous
cosmological relics such as the moduli and gravitinos are all diluted away \cite{thermal,axion-thermal}.

The scalar boson with a mass $m_h\simeq 125$~GeV, which was recently discovered in the LHC experiments,
has been found to behave like the SM Higgs boson \cite{ATLAS-Higgs,CMS-Higgs}.
On the other hand, a SM-like Higgs boson at 125~GeV in the minimal supersymmetric standard model (MSSM)
requires that stops have either a heavy mass in multi-TeV range or maximal LR-mixing, which would cause
a fine-tuning worse than 1~\% in the electroweak symmetry breaking.
This fine-tuning can be ameliorated in the next-to-minimal supersymmetric standard model (NMSSM) involving
a singlet superfield $S$ with the superpotential coupling $\lambda SH_uH_d$.
In the NMSSM, the SM-like Higgs boson $h$ gains an additional tree-level mass from the $F$-term scalar
potential $\lambda^2 |H_uH_d|^2$, or from scalar mixing if the singlet scalar $s$ is lighter than $h$.
This makes it possible to have $m_h\simeq 125$~GeV even when stops are  relatively light and stop mixings are small,
and therefore reduces the amount of fine-tuning required for the electroweak symmetry breaking
\cite{Review-NMSSM,fine-tuning}.
Furthermore the Higgs and neutralino sector of the NMSSM have a richer structure than the MSSM.
If all the Higgs bosons in the NMSSM have masses in sub-TeV range, e.g. below 500~GeV,
there can be sizable mixings among the three CP-even Higgs bosons, leading to interesting
phenomenological consequences as discussed in \cite{fine-tuning,diphoton-enhancement-nmssm,Diphoton_NMSSM,nmssm-higgs}.

It is well known that a PQ-symmetry spontaneously broken at $v_{\rm PQ}\sim \sqrt{m_{\rm soft}M_{Pl}}$
can explain why the doublet-higgsino mass $\mu$ in the MSSM is comparable to $m_{\rm soft}$
\cite{axion-scale,axion-thermal,Jeong:2011xu,mu-problem}.
Similarly, if the singlet superfield $S$ is PQ-charged, the low energy mass parameters of $S$ in
the effective superpotential of the PQ-invariant NMSSM are induced by a spontaneous breakdown
of the PQ symmetry, and so can have a value comparable to $m_{\rm soft}$, while the singlet cubic
coupling is always negligible \cite{PQ-NMSSM,Bae:2012am}.
In this paper, we wish to examine the Higgs phenomenology in such a PQ-invariant NMSSM while
focusing on the phenomenological consequences of scalar mixing.
The Higgs boson masses and mixing angles in the neutral CP-even Higgs sector crucially depend
on the coupling $\lambda$ and the doublet-higgsino mass $\mu$.
As a result, scalar mixing is constrained not only by the observed mass and signal strengths
of the SM-like Higgs boson, but also by the perturbativity bound on $\lambda$ and the LEP bound on the chargino mass.
We will examine first the constraints on scalar mixing in the context of the general NMSSM,
and then consider a specific minimal PQ-invariant NMSSM which is further constrained by the presence of a light
singlino-like neutralino.

If the singlet-like Higgs boson $s$ has a mass near the weak scale, it can have a large mixing
with the SM-like Higgs boson $h$.
We identify the parameter region of the sizable singlet-doublet mixing that is compatible with
all the LHC and LEP data available at present, as well as with the perturbativity bound
on $\lambda$ and a stop mass between 600~GeV and a few TeV.
We explore also the possibility that the $2\sigma$ excess of the LEP $Zb\bar b$ events at
$m_{b\bar b}\simeq 98$~GeV is explained by $e^+e^-\rightarrow Zs\rightarrow Zb\bar b$ within
the framework of the minimal PQ-NMSSM.\footnote{Such a possibility for the conventional $Z_3$-invariant
NMSSM has been examined recently in Ref.~\cite{Belanger:2012}.}
We then find that it requires low $\tan\beta$ smaller than about 2, a light doublet-higgsino mass around
the weak scale, and stop masses around or below 1~TeV.
For the case with $m_s>m_h$, it is found that $s$ decays dominantly into a neutralino pair in
most of the viable parameter region, which would make its detection at collider experiments difficult.
We examine also the signal rates of the SM-like Higgs boson in the $b\bar b$ ($\tau\bar\tau$) and
di-photon channels over the phenomenologically viable parameter region which gives the signal
rate of the $WW/ZZ$ channel close to the SM value.

This paper is organized as follows. In section \ref{sec:Higgs-NMSSM}, we discuss how the constraints
on $\lambda$ and $\mu$, and the observed properties of the SM-like Higgs boson translate into
the constraints on the scalar mixing angles.
In section \ref{sec:PQ-NMMS}, we discuss some generic features of the PQ-invariant NMSSM,
and present a specific model which is considered to be a minimal PQ-invariant NMSSM with
$v_{\rm PQ}\sim \sqrt{m_{\rm soft}M_{Pl}}$.
The neutralino sector of the minimal PQ-NMSSM is also discussed with a focus on the additional
constraints arising due to a light singlino-like neutralino in the model.
In section \ref{sec:Higgs-phenomenology}, we apply the results of the section \ref{sec:Higgs-NMSSM}
to the Higgs phenomenology in the minimal PQ-NMSSM.
We present first the results that hold in the general NMSSM, and then impose additional constraints
specific to the minimal PQ-NMSSM.
Section \ref{sec:conclusions} is the conclusions.

\section{Constraints on Higgs mixing in the NMSSM}\label{sec:Higgs-NMSSM}

In this section, we briefly discuss phenomenological consequences of Higgs mixing in the general NMSSM and the resultant constraints on the model.
Let us begin with the Higgs sector superpotential of the general NMSSM, which is given by
\bea
\lambda S H_uH_d + f(S),
\eea
in an appropriate basis of the singlet superfield $S$.
The first term is responsible for the higgsino mass parameter $\mu$ and Higgs bilinear
coupling $B\mu$:
\bea
\mu &=& \lambda\langle S \rangle,
\nonumber \\
B\mu &=& \lambda \left( A_\lambda \langle S \rangle + \langle \partial_S f \rangle\right),
\eea
where $A_\lambda$ is the soft SUSY breaking parameter for the superpotential
term $SH_uH_d$.
There is one combination $\hat h$ of CP-even neutral Higgs bosons which corresponds to the
fluctuation of ${\rm Re}(H^0_u)$ and ${\rm Re}(H^0_d)$ in the vacuum value direction,
and therefore behaves like the SM Higgs boson in the limit when the other Higgs bosons are decoupled.
In the NMSSM, it generally mixes with the other CP-even neutral Higgs bosons, and the SM-like
Higgs boson in the mass-eigenstate is given by
\bea
h = c_{\theta_1}c_{\theta_2} \hat h -s_{\theta_1}\hat H -c_{\theta_1}s_{\theta_2} \hat s,
\eea
with $c_{\theta_i}= \cos\theta_i$ and $s_{\theta_i}=\sin\theta_i$ for the mixing angles $\theta_i$
defined in appendix \ref{appendix},
where $\hat H$ is the fluctuation of ${\rm Re}(H^0_u)$ and ${\rm Re}(H^0_d)$ orthogonal
to $\hat h$, and $\hat s$ is the CP-even fluctuation of the singlet scalar.

Around the weak scale, the SM-like Higgs boson interacts with the SM particles through
the terms\footnote{ This should be understood as an 1PI effective Lagrangian including
quantum corrections for the SM-like Higgs boson near the mass-shell.}
\cite{Carmi:2012in},
\bea
\label{Higgs-coupling}
{\cal L} &=& C_V
\frac{\sqrt2m^2_W}{v} h W^+_\mu W^-_\mu
+ C_V \frac{m^2_Z}{\sqrt2 v} h Z_\mu Z_\mu
- C_f \frac{m_\psi}{\sqrt2 v}h \bar f f
\nonumber \\
&&
+\, C_g \frac{\alpha_s}{12\sqrt2\pi v} h G^a_{\mu\nu} G^a_{\mu\nu}
+ C_\gamma \frac{\alpha}{\sqrt2\pi v} h A_{\mu\nu} A_{\mu\nu},
\eea
where  $f$ denote the SM fermions, and $v\simeq 174 $ GeV is the Higgs vacuum expectation value.
At tree level, the Higgs couplings to the massive SM particles are determined
by the mixing angles as
\bea
C_V &=& c_{\theta_1}c_{\theta_2},
\nonumber \\
C_t &=& c_{\theta_1}c_{\theta_2}+s_{\theta_1}\cot\beta,
\nonumber \\
C_b &=& C_\tau = c_{\theta_1}c_{\theta_2}-s_{\theta_1}\tan\beta.
\eea
On the other hand, the Higgs couplings to massless gluons and photons are radiatively induced.
The dominant contribution comes from the $W$-boson and top quark loops:
\bea
C_g &\simeq& 1.03 C_t -0.06 C_b + \delta C_g,
\nonumber \\
C_\gamma &\simeq& 0.23 C_t -1.04C_V + \delta C_\gamma,
\eea
including superparticle loop contributions $\delta C_g$ and $\delta C_\gamma$,
where $\delta C_g$ can be sizable if stops are below 1~TeV, and $\delta C_\gamma$
becomes important if there are light charged-superparticles around the weak scale.
Using the above relations, one can estimate the signal rate of the SM-like Higgs boson $h$
at the LHC in the presence of scalar mixing.
The signal rate in the $WW/ZZ$ channel normalized by the SM value is given by
\bea
\label{RVV}
R^{VV}_h \simeq
\frac{(0.94 C^2_g + 0.12 C^2_V)C^2_V}{0.64 C^2_b+0.24C^2_V+0.12 C^2_t},
\eea
where we have used the well-known production and decay properties of the SM Higgs boson
under the assumption that the Higgs decay rate into non-SM particles is negligible.
To see the effect of scalar mixing, it is convenient to factor the signal rate into $WW/ZZ$ as
\bea
\label{relation-for-Rvv}
R^{VV}_h \simeq \left(1 + 2 \frac{\delta C_g}{C_g} \right) R^{VV}_h|_0,
\eea
where $R^{VV}_h|_0$ is the signal rate for $\delta C_g=0$, i.e. in the limit
that all the colored superparticles are  heavy.
It is important to note that $R^{VV}_h|_0$ depends only on $\theta_1$, $\theta_2$,
and $\tan\beta$.
In addition, because the effect of colored superparticles is to modify the Higgs production
rate in the gluon fusion process, the ratio $R^{ii}_h/R^{VV}_h$ for each channel is insensitive
to the correction $\delta C_g$.
For other channels, we find
\bea
R^{bb}_h &=& R^{\tau\tau}_h = \frac{C^2_b}{C^2_V} R^{VV}_h,
\nonumber \\
R^{\gamma\gamma}_h &\simeq&
\frac{1.52 C^2_\gamma}{C^2_V} R^{VV}_h,
\eea
where $R^{ii}_h=1$ in the limit of vanishing mixing angles
and decoupled superparticles.

In the NMSSM, the Higgs quartic coupling receives an additional tree-level contribution
proportional to $\lambda^2$, and consequently $\hat h$ obtains a mass according to
\bea
\label{mhhat}
m^2_{\hat h} &=& m^2_0 + (\lambda^2 v^2-m^2_Z)\sin^2 2\beta,
\eea
where $m_0$ corresponds to the SM-like Higgs boson mass at large $\tan\beta$ in the decoupling
limit of MSSM, including the well-known radiative correction from top and stop
loops \cite{Okada:1990vk}:
\bea
m^2_0 &=&
m^2_Z
+ \frac{3m^4_t}{4\pi^2v^2} \ln\left(\frac{m^2_{\tilde t}}{m^2_t}\right)
+ \frac{3m^4_t}{4\pi^2v^2}\left( X^2_t - \frac{1}{12}X^4_t\right)
+ \cdots,
\eea
for the stop mass $m_{\tilde t}$ and the stop mixing parameter
$X_t=(A_t-\mu\cot\beta)/m_{\tilde t}$.
It is straightforward to see that the mass of the SM-like Higgs boson in the NMSSM reads
\bea
\label{Higgs-mass}
m^2_h &=& m^2_{\hat h}
- \frac{(s_{\theta_2}s_{\theta_3}-s_{\theta_1}c_{\theta_2}c_{\theta_3})^2}
{c^2_{\theta_1}c^2_{\theta_2}}
(m^2_H-m^2_{\hat h})
- \frac{(s_{\theta_2}c_{\theta_3}+s_{\theta_1}c_{\theta_2}s_{\theta_3})^2}
{c^2_{\theta_1}c^2_{\theta_2}}
(m^2_s-m^2_{\hat h}),
\eea
where the last two terms are due to scalar mixing.
Note that the mixing with singlet scalar increases $m_h^2$ if the singlet-like Higgs boson $s$
is lighter than the SM-like Higgs boson $h$ \cite{arXiv:0712.2903, arXiv:1108.4338, PQ-NMSSM}.

In the presence of scalar mixing, the singlet-like Higgs boson $s$ also interacts with the SM particles
via the doublet components.
Those interactions are obtained from (\ref{Higgs-coupling}) by replacing $C_i$ with the effective
couplings
\bea
C^{s}_V &=& s_{\theta_2} c_{\theta_3} + s_{\theta_1} c_{\theta_2} s_{\theta_3},
\nonumber \\
C^{s}_t &=& s_{\theta_2} c_{\theta_3} + s_{\theta_1} c_{\theta_2} s_{\theta_3}
- c_{\theta_1} s_{\theta_3} \cot\beta,
\nonumber \\
C^{s}_b &=&  C^{s}_\tau =
s_{\theta_2} c_{\theta_3} + s_{\theta_1} c_{\theta_2} s_{\theta_3}
+ c_{\theta_1} s_{\theta_3} \tan\beta,
\eea
at tree-level, and the coupling to gluons and photons are radiatively generated depending on the singlet
mass $m_s$.

Let us examine how the SM-like Higgs boson in the NMSSM can be arranged to be consistent with the LHC data.
The most important constraints come from the mass and signal rates for the various Higgs decay channels observed
at the LHC.
In particular, the signal rate for the $WW/ZZ$ channel should be close to the SM value,
\bea
R^{VV}_h \approx 1,
\eea
which does not necessarily imply that the $h$-$s$ mixing angle $\theta_2$ should be small.
Keeping in mind that the Higgs coupling to gluons can receive a non-negligible correction from
relatively light stops, we impose the condition
\bea
R^{VV}_h|_0 \simeq 1,
\eea
to account for the observed Higgs signal rate in $WW/ZZ$.
This  is the case when the mixing angles obey the relation \cite{Diphoton_NMSSM},
\bea
\label{RVV-soln}
\theta_1 &\approx& \frac{\tan\beta}{1.4\tan^2\beta+1.7} \sin^2\theta_2.
\eea
Here we have used that $R^{VV}_h|_0$ is determined only by $\theta_1$, $\theta_2$,
and $\tan\beta$.
For such Higgs mixing, the signal rates for the fermionic ($b\bar b$ or $\tau\bar\tau$) and
di-photon channel are estimated to be
\bea
R^{bb}_h &=& R^{\tau\tau}_h \approx
\left(
1 -  \theta_1 \tan\beta \right)^2 R^{VV}_h,
\nonumber \\
R^{\gamma\gamma}_h &\approx& \left(
1 -0.28 \theta_1 \cot\beta - 1.23\, \delta C_\gamma \right)^2 R^{VV}_h,
\eea
with $R^{VV}_h\approx 1$.
This shows that the signal rates for the $b\bar b$ and $\tau\bar\tau$ channel are reduced below
the SM prediction as a result of scalar mixing at tree level.
The di-photon rate is less affected by scalar mixing.
However, in the presence of sizable $\theta_2$ and light charged-higgsinos, it can significantly
deviate from the SM value due to the chargino-loop contribution to $\delta C_\gamma$,
which is given by
\cite{Carmi:2012in}
\bea
\delta C_\gamma|_{\tilde H^\pm} &\approx&
-0.17\frac{\lambda v}{|\mu|} \tan\theta_2.
\eea
Note that the charged-higgsino loop can either enhance or reduce the di-photon rate, depending
on the sign of $\theta_2$.

In the NMSSM, for a given value of $\tan\beta$, the off-diagonal components of the mass matrix of
$(\hat h, \hat H, \hat s)$ are determined by three parameters $\{\lambda, \mu, \Lambda\}$
(see appendix \ref{appendix}), where
\bea
\Lambda = A_\lambda + \langle \partial^2_S f  \label{Lambda}\rangle
\eea
is independent from the effective Higgs bilinear coupling $B\mu$.
These parameters can be expressed in terms of the mixing angles $\theta_i$ and the mass
eigenvalues $m_h$, $m_H$ and $m_s$.
In particular, $\lambda$ and $\mu$ are given by
\bea
\label{Lagrangian_p}
\lambda^2 v^2 &=& m^2_Z
+ \frac{1}{\sin4\beta}
\Big( (m^2_H - m^2_s) s_{\theta_2} s_{2\theta_3}
+ 2(m^2_h - m^2_H c^2_{\theta_3} - m^2_s s^2_{\theta_3})
s_{\theta_1}c_{\theta_2}
\Big) c_{\theta_1},
\nonumber \\
\lambda v \mu &=&
-\frac{1}{4} m^2_h c^2_{\theta_1} s_{2\theta_2}
-\frac{1}{4}(m^2_H-m^2_s) s_{\theta_1} c_{2\theta_2}s_{2\theta_3}
\nonumber \\
&& +\, \frac{1}{4} \Big(
(m^2_H-m^2_s s^2_{\theta_1}) s^2_{\theta_3}
- (m^2_H s^2_{\theta_1} - m^2_s) c^2_{\theta_3}
\Big)s_{2\theta_2}
\nonumber \\
&& -\, \frac{\tan2\beta}{4} \Big(
(m^2_H-m^2_s) c_{\theta_2} s_{2\theta_3}
 - 2(m^2_h-m^2_H c^2_{\theta_3} -m^2_s s^2_{\theta_3})
s_{\theta_1}s_{\theta_2}
\Big)c_{\theta_1}.
\eea
On the other hand, the coupling $\lambda$ is constrained to be less than about 0.7 at the weak scale,
if one wishes to maintain the model to be perturbative up to the GUT scale \cite{Miller:2003ay},
while the LEP bound on the chargino mass requires $|\mu|$ to be larger than about 100~GeV \cite{PDG}.
These constraints on $\lambda$ and $\mu$ can be translated into those on the mixing angles and mass
eigenvalues through the above relations.

Finally, $m_0$ cannot take an arbitrary value, and thus Higgs mixing is constrained by the requirement
$m_h\simeq 125$ GeV through the relation (\ref{Higgs-mass}).
For instance, the stop searches at the LHC suggest that the stop is heavier than about 600~GeV \cite{ATLAS-stop,CMS-stop}, implying $m_0\gtrsim 105$ GeV.
One may avoid this stop mass bound by considering the case where the stop mass is smaller than
the sum of the top quark mass and the lightest neutralino mass.
On the other hand, fine-tuning for the electroweak symmetry breaking becomes more
severe for heavier stop masses,
so the naturalness principle favors $m_0$ to be as small as possible.
We therefore assume $m_0$ to be in the range,
\bea
105\,{\rm GeV}\,\lesssim\,m_0 \,\lesssim\, 120\,{\rm GeV},
\label{m_0r}
\eea
which amounts to assuming that stops are not significantly heavier than 1~TeV.
Note that $\delta C_g$ receives the dominant contribution from stop loops
\cite{Arvanitaki:2012, Blum:2013},
\bea
{\delta C_g} \approx \frac{1}{4}\frac{m_t^2}{m_{\tilde{t}}^2}(2-X_t^2)C_t ,
\eea
which can be sizable for stop masses of our interest.
This correction to the Higgs coupling to gluons modifies the Higgs production rate in the gluon fusion, and
enhances the Higgs signal rates for the stop mixing parameter $X_t < \sqrt{2}$.
For instance, taking $X_t=0$, one finds that the Higgs signal rate in each channel
 increases universally by about 8~\% for the stop mass around 600~GeV, and by less than about 3~\% for the stop mass
heavier than 1~TeV.
Here the ratio $R^{ii}_h/R^{VV}_h$ remains almost the same.
The stop contribution to the Higgs-photon coupling $C_\gamma$ is below 1~\% even for the stop mass
around 600~GeV.

We close this section by summarizing the conditions yielding constraints on the Higgs mixing
in the general NMSSM.
These include $(a)$ the mass of the SM-like Higgs boson, $m_h \simeq 125$~GeV, $(b)$ the Higgs
signal rates, in particular $R^{VV}_h \approx 1$, $(c)$ the perturbativity bound $\lambda\lesssim 0.7$,
and $(d)$ the LEP bound on the chargino mass, implying $|\mu|\gtrsim 100$~GeV.

\section{Peccei-Quinn invariant NMSSM}\label{sec:PQ-NMMS}

In this section we discuss the generic low energy limit of the PQ-invariant NMSSM, and present
a specific model considered to be a minimal PQ-NMSSM.
As we will see, a key feature of the minimal PQ-NMSSM is the presence of a light singlino-like neutralino,
with which the model is severely constrained by the Higgs invisible decay and the LEP bound on neutralino
productions.

\subsection{Low energy limit of the generic PQ-NMSSM}

At energy scales below the PQ-breaking scale $v_{\rm PQ}$, the PQ-NMSSM can be described
by a low energy effective theory with a non-linear $U(1)_{\rm PQ}$ symmetry,
under which the NMSSM Higgs superfields and the axion superfield $A$ transform as
\bea
S &\to& e^{i\alpha} S,
\nonumber \\
H_uH_d &\to& e^{-i\alpha} H_uH_d,
\nonumber \\
A &\to& A + iv_{\rm PQ}\alpha.
\eea
Throughout this paper, we assume that the PQ-breaking scale is generated by  competition
between SUSY breaking effect and Planck scale suppressed effect, so that
\bea
v_{\rm PQ}\sim \sqrt{m_{\rm soft}M_{Pl}}.
\nonumber
\eea
Here the axion superfield $A$ is composed of a pseudo-scalar axion $a$ solving the strong CP problem,
its scalar partner saxion $\rho$, and the fermionic partner axino $\tilde a$:
\bea
A =
\frac{1}{\sqrt2} (\rho + i a ) + \sqrt2 \theta \tilde a
+ \theta^2 F^A.
\eea
The PQ-invariant K\"ahler potential and superpotential below $v_{\rm PQ}$ are generically given by
\bea
\label{PQ-NMSSM-action}
K &=& K_0(A+A^*)
+ \sum_i Z_i(A+A^*)|\Phi_i|^2 + \Delta K,
\nonumber \\
W &=& \left(\mbox{MSSM Yukawa terms}\right) + \lambda S H_uH_d + \Delta W,
\eea
in which $\Phi_i$ denote the NMSSM chiral superfields.
Here $\Delta K$ and $\Delta W$ stand for the terms induced by a spontaneous breakdown
of the PQ symmetry,
\bea
\Delta K &=&\tilde \mu_1\, e^{A^*/v_{\rm PQ}}S + \kappa_1\, e^{2A^*/v_{\rm PQ}}S^2
+ \kappa_2\, e^{-A^*/v_{\rm PQ}}H_uH_d + \cdots +{\rm h.c.},
\nonumber \\
\Delta W &=& \tilde \mu_2^2\,e^{-A/v_{\rm PQ}}S +\tilde \mu_3\, e^{-2A/v_{\rm PQ}}S^2
+ \tilde \mu_4\, e^{A/v_{\rm PQ}} H_uH_d + \kappa_3\, e^{-3A/v_{\rm PQ}}S^3 + \cdots,
\eea
where the ellipses denote higher dimensional terms, and
\bea
\kappa_j \,\lesssim\,
{\cal O}\Big(\left({v_{\rm PQ}}/{M_{Pl}}\right)^{k_j}\Big), \quad
\frac{\tilde \mu_j}{v_{\rm PQ}} \,\lesssim\,
{\cal O}\Big(\left({v_{\rm PQ}}/{M_{Pl}}\right)^{n_j}\Big),
\eea
for model-dependent non-negative integers $k_j$ and $n_j$.

Including the effects of soft SUSY-breaking, the vacuum value of the axion superfield can be
determined to be\footnote{The axion vacuum value is not determined by SUSY breaking effects,
but fixed by the low energy QCD dynamics at a value solving the strong CP problem.}
\bea
\frac{\langle A \rangle}{v_{\rm PQ}} &=& \xi_1 +\xi_2 m_{\rm soft} \theta^2,
\eea
where $\xi_{1,2} ={\cal O}(1)$ in general.
To examine the particle physics phenomenology at scales below $v_{\rm PQ}$, it is convenient
to replace the axion superfield with its vacuum expectation value.
After this replacement, one can make an appropriate field redefinition
\bea
S\, \rightarrow \, S + \mu_0 + b_0 \theta^2, \label{redefinition} \eea
together with a K\"ahler transformation
\bea
K_{\rm eff} \,\rightarrow \, K_{\rm eff} - (\theta^2 \Omega + {\rm h.c.}),
\quad
W_{\rm eff}\, \rightarrow W_{\rm eff} + \Omega,
\label{kahlertrans}
\eea
to arrive at the following form of the effective K\"ahler potential and superpotential:
\bea
K_{\rm eff} &=& \sum_i \left(1- m^2_i\theta^2\bar\theta^2\right) |\Phi_i|^2,
\nonumber  \\
W_{\rm eff} &=& \left(\mbox{MSSM Yukawa and $A$-terms}\right)
+ \lambda (1+A_\lambda\theta^2) SH_uH_d
\nonumber \\
&&
+\,
\mu_1^2 (1+B_1\theta^2) S + \frac{1}{2} \mu_2 (1+B_2\theta^2)S^2
+ \frac{1}{3} \kappa (1+A_\kappa \theta^2)S^3,
\eea
where
\bea
m_i &\sim& A_{\lambda,\kappa}  \,\sim\,  B_{1,2} \,\sim\, m_{\rm soft},
\nonumber \\
\mu_{1,2} &\sim&   m_{\rm soft} \left(\frac{v_{\rm PQ}}{M_{Pl}}\right)^{n_{1,2}}
\quad (n_{1,2} \geq 0),
\nonumber \\
\kappa &\sim&
\left(\frac{v_{\rm PQ}}{M_{Pl}}\right)^{n_0}
\,{\rm or}\,
\left(\frac{m_{\rm soft}}{M_{Pl}}\right)
\left(\frac{v_{\rm PQ}}{M_{Pl}}\right)^{n_0}
\quad (n_0\geq 1),
\eea
with a PQ scale given by  $v_{\rm PQ}\sim \sqrt{m_{\rm soft}M_{Pl}}$.
A simple generic feature of the PQ-NMSSM is that the singlet cubic coupling $\kappa$ is always negligible,
\bea
\kappa &\lesssim& {\cal O}(v_{PQ}/M_{Pl}) \,\sim\,  10^{-7}-10^{-8},
\eea
while the singlet mass parameters $\mu_{1,2}$ can be either of the order of $m_{\rm soft}$ or
negligibly small compared to $m_{\rm soft}$, depending on the relative charge between $S$
and PQ-breaking fields.

\subsection{A minimal PQ-NMSSM}\label{sec:PQ-neutralino-sector}

In this subsection, we present a specific model  which is considered to be a minimal PQ-invariant NMSSM,
and discuss the neutralino sector of the model.
At high scales above $v_{\rm PQ}$, but below the Planck scale $M_{Pl}$, the model includes the PQ-breaking
superfields $X$ and $Y$, as well as the NMSSM Higgs superfields $S$, $H_u$ and $H_d$, with
the following PQ-charges:
\bea
(S, H_uH_d, X, Y) = (1,-1, \,\frac{1}{2}\,, -\frac{1}{6}).\eea
The model can include also exotic gauge-charged matter superfields $\Psi_I,\Psi^c_I$ ($I=1,2$), which are
vector-like under the SM gauge group, e.g. ${\bf 5}+\bar{\bf 5}$ of SU$(5)$, and carry a PQ-charge which
allows renormalizable Yukawa couplings to $X$ or $Y$, e.g.
\bea
(\Psi_1\Psi_1^c, \Psi_2\Psi_2^c)=(-\frac{1}{2},\,\frac{1}{6}\,).
\eea
Then the most general PQ-invariant K\"ahler potential and superpotential are written as
\bea
K &=& \sum_i|\Phi_i|^2 + \frac{1}{M_{Pl}}\left(y_1X^2S^*+{\rm h.c.}\right)+ \cdots,
\nonumber \\
W &=& \lambda S H_uH_d + \lambda^\prime X\Psi_1\Psi_1^c + \lambda^{\prime\prime}Y\Psi_2\Psi_2^c
+ \frac{1}{M_{Pl}}\left(y_2 X^2H_uH_d + y_3 XY^3\right) + \cdots,
\eea
where the ellipses denote higher dimensional operators suppressed by higher powers of $1/M_{Pl}$.
Including soft SUSY breaking terms, the scalar potential of the PQ-breaking fields takes the form
\bea
V &=& m_X^2|X|^2+m_Y^2|Y|^2+ \left( \frac{y_3 A_3}{M_{Pl}} XY^3+{\rm h.c.}\right)
+ \frac{y_3^2}{M_{Pl}^2}|Y|^6 + \cdots.
\eea
Assuming $m_Y^2 <0$ and $m_X^2>0$ around the renormalization point
$\sim\sqrt{m_{\rm soft}M_{Pl}}$,
which can be a consequence of either a $D$-term induced soft SUSY breaking or
the radiative correction due to a large Yukawa coupling  $\lambda^{\prime\prime}$,
one finds
\bea
\langle X \rangle \,\sim\,  \langle Y\rangle \,\sim\, \sqrt{m_{\rm soft}M_{Pl}},
\quad
\frac{F^X}{X}\,\sim\, \frac{F^Y}{Y}\,\sim m_{\rm soft}, \eea
assuming that
\bea
|m_X| \sim |m_Y| \sim |A_3| \sim m_{\rm soft}, \quad y_{3}={\cal O}(1).
\nonumber
\eea
Now we can replace the PQ-breaking superfields $X$ and $Y$ with their vacuum expectation
values while including the soft SUSY-breaking terms explicitly.
Making further a field redefinition of (\ref{redefinition}) and a K\"ahler transformation of
(\ref{kahlertrans}), we find that the resulting low energy effective theory takes the form
\bea
K_{\rm eff} &=& \sum_i \left(1- m_i^2\theta^2\bar\theta^2\right) |\Phi_i|^2,
\nonumber  \\
W_{\rm eff} &=&  \left(\mbox{MSSM Yukawa and $A$-terms}\right)
+ \lambda (1+A_\lambda\theta^2) SH_uH_d + \mu_1^2 (1+B_1\theta^2) S,
\eea
where the quadratic and cubic terms of $S$ in $W_{\rm eff}$ are omitted since their coefficients
are negligibly small:
\bea
\mu_2 \,\sim\, m_{\rm soft}\left(\frac{v_{\rm PQ}}{M_{Pl}}\right)^4, \quad \kappa\,\sim\,
\left(\frac{m_{\rm soft}}{M_{Pl}}\right)\left(\frac{v_{\rm PQ}}{M_{Pl}}\right)^6.
\nonumber \eea

Although the effective superpotential of this minimal PQ-NMSSM takes a simple form,\footnote{
Note that $W_{\rm eff}$ of the minimal PQ-NMSSM is the same as that of the nMSSM,
which has been proposed in Refs.~\cite{nMSSM1,nMSSM2,nMSSM3} in a different context.
See also Refs.~\cite{DM-nMSSM,PH-nMSSM}.
}
the Higgs sector of the model is not distinctive as the Higgs mixing parameter
$\Lambda= A_\lambda +\langle \partial_S^2 W_{\rm eff}\rangle=A_\lambda$ is still independent from $B\mu$.
On the other hand, the neutralino sector of the model is quite distinctive since $\partial_S^2 W_{\rm eff}=0$,
and therefore the singlino gains a mass only through the mixing with other neutralinos:
\bea
-{\cal L}_{\chi^0_i} = \mu \tilde H^0_u \tilde H^0_d
+ \lambda v \cos\beta \tilde H^0_u \tilde S
+ \lambda v \sin\beta \tilde H^0_d \tilde S
+ \cdots,
\eea
where the ellipsis denotes the gaugino mass and gaugino-higgsino mixing terms.
It can be shown that the lightest neutralino,
\bea
\chi^0_1 &=& N_{11} \tilde B + N_{12} \tilde W^0
+ N_{13}\tilde H^0_d + N_{14} \tilde H^0_u + N_{15} \tilde S,
\nonumber
\eea
has a mass lighter than $\lambda v \cos\beta$ in the limit when the mixing with gauginos
is ignored \cite{Bae:2012am}.
To see qualitatively the properties of the lightest neutralino in the minimal PQ-NMSSM,
one can take the limit of $\mu \gg \lambda v $ and the gaugino masses $M_i \gg v$.
Then the neutralino mixing coefficients are found to be
\bea
N_{13} &=& - \frac{\lambda v\cos\beta}{\mu} + {\cal O}\left(\left({\lambda v}/{\mu}\right)^2\right),
\nonumber \\
N_{14} &=& -\frac{\lambda v\sin\beta}{\mu} + {\cal O}\left((\lambda v/\mu)^2\right),
\nonumber \\
N_{15} &=& 1 -\frac{\lambda^2 v^2}{2\mu^2} + {\cal O}\left((\lambda v/\mu)^3\right),
\eea
while the mass eigenvalue is given by
\bea
\label{neutralino-mass}
m_{\chi^0_1} &=& \frac{\lambda^2 v^2\sin 2\beta}{\mu} \left(1+{\cal O}\left((\lambda v/\mu)^2\right)\right).
\label{neut_mass}
\eea
The gaugino components in $\chi^0_1$ are generally small because they are further suppressed
by $v/M_i \ll 1$:
\bea
|N_{1i}| = g_i\frac{\lambda v^2 \cos2\beta}{\sqrt2 \mu M_i}
\left(1 + {\cal O}\left(\lambda v/\mu\right)\right),
\eea
for $i=1,2$ with $M_i$ being the corresponding gaugino mass.

There are important constraints on the minimal PQ-NMSSM associated with the small mass
of the lightest neutralino.
One is from the LEP bound on the neutralino production via the $Z$-boson exchange \cite{OPAL}:
\bea
\sigma(e^+e^-\to \chi^0_2\chi^0_1)\times {\rm Br}(\chi^0_2 \to q\bar q \chi^0_1)
\,\lesssim\, 100\,{\rm fb},
\eea
which applies for $m_{\chi^0_2}+m_{\chi^0_1}<208$~GeV and $m_{\chi^0_1}>60$~GeV.
This puts an upper bound on the $Z$-boson coupling to $\chi^0_2\chi^0_1$.
In addition, the global fit analysis excludes an invisible decay of the SM-like Higgs boson
with a branching ratio greater than 0.38 at 95\% confidence level,
if one allows its couplings to the SM particles to deviate from the SM values
\cite{Giardino:2013bma,Belanger:2013}:
\bea
{\rm Br}(h\to \chi^0_1\chi^0_1) < 0.38.
\eea
In the NMSSM, the Higgs coupling for this process is given by
\bea
y_{h\chi^0_1\chi^0_1} =  \frac{\sqrt2\lambda^2 v\sin 2\beta}{\mu}
\left(1+
{\cal O}\left({\lambda v}/{\mu}\right) \right),
\eea
which can have a sizable value in the limit of low $\tan\beta$, large $\lambda$,
and light $\mu$.
To avoid a dangerous Higgs invisible decay when the Higgs coupling $y_{h\chi^0_1\chi^0_1}$
is sizable, one needs $2m_{\chi^0_1}>m_h$ so that the process is kinematically forbidden.
On the other hand, if $2m_{\chi^0_1}<m_h$, one needs to adjust the model to suppress
$y_{h\chi^0_1\chi^0_1}$.
However, with small $y_{h\chi^0_1\chi^0_1}$, it is difficult to have a sizable NMSSM contribution
to the tree level mass of the SM-like Higgs boson, which is the feature that we like to keep
to avoid too severe fine-tuning of the model.
Note that, since we are assuming $m_0 \lesssim 120 $~GeV, a sizable NMSSM contribution is required
to get $m_h \simeq 125$~GeV.
Actually, as we shall see in the next section, this makes it difficult to suppress the branching
fraction for the Higgs invisible decay below 0.38 in most of the parameter space of our interest
once the mode is kinematically open.
We therefore require $2m_{\chi^0_1}>m_h$ to prohibit the decay process
$h\rightarrow \chi^0_1\chi^0_1$.

\section{Higgs phenomenology of the PQ-NMSSM}\label{sec:Higgs-phenomenology}

The SM-like Higgs boson observed at the LHC can be accommodated in the NMSSM while satisfying
the constraints on scalar mixing discussed in sec.~\ref{sec:Higgs-NMSSM}.
We will first examine how large the mixing between the SM-like Higgs boson and the singlet-like Higgs boson
is allowed in the general NMSSM, and then
move on to the minimal PQ-NMSSM where the mixing is further constrained due to a light singlino-like neutralino.
As we will see, a SM-like Higgs boson with sizable singlet component can be compatible with
all the LHC and LEP results available at present.
In such case, the singlet scalar is expected to be around the weak scale since otherwise
a sizable singlet-doublet mixing would make it difficult to explain the observed SM-like Higgs boson
mass $m_h\simeq 125$~GeV.
The singlet-like Higgs boson $s$ can be lighter or heavier than the SM-like Higgs boson $h$.
For the former case, we will focus on the possibility that the $2\sigma$ excess of the LEP $Zb\bar b$
events at $m_{b\bar b}\simeq 98$~GeV is explained by $s$ with $m_s \simeq 98$~GeV.

Let us briefly explain how we explore the effect of scalar mixing.
The relations (\ref{Higgs-mass}) and (\ref{Lagrangian_p}) will be used to express $\{\lambda, \mu, m_0\}$
in terms of $\{\theta_i,\tan\beta,m_s,m_H\}$ with $m_h\simeq 125$~GeV.
We also require that the scalar mixing angles obey the relation (\ref{RVV-soln}) in order for the Higgs
signal rate for the $WW/ZZ$ channel to be close to the SM prediction.
Then, taking some benchmark values of the Higgs boson masses $m_s$ and $m_H$, one can see how the model
parameters $\{\lambda, \mu, m_0\}$ change on the parameter plane $(\theta_2,\tan\beta)$ for a given
value of $\theta_3$.
In other words, it is possible to figure out which region in the $(\theta_2,\tan\beta)$ space is allowed
by the constraints on $\{\lambda, \mu, m_0\}$.
Over the allowed region, we will examine the signal rates of the SM-like Higgs boson
for the $f\bar f$ ($f=b,\tau$) and $\gamma\gamma$ channel, and also the properties of the singlet-like
Higgs boson.

\subsection{Singlet-like Higgs boson at 98  GeV}

If the singlet-like Higgs boson $s$ is lighter than 114~GeV, scalar mixing is constrained
not only by the LHC results, but also by the LEP search of the Higgs boson.
In regard to this possibility, a particularly interesting LEP result is the $2\sigma$ excess
of $Zb\bar b$ around $m_{b\bar b}\simeq 98$~GeV \cite{LEP98GeV}.
This may indicate a light singlet-like Higgs boson with $m_s\simeq 98$~GeV, which can have a sizable
coupling to the $Z$ boson through the mixing with the SM-like Higgs boson:
\bea
C^s_V \approx \theta_2.
\eea
Let $R_s^{Zb\bar b}$ be the signal strength for $e^+e^-\to Zs \to Zb\bar b$ normalized
by the signal rate for the SM Higgs boson having the same mass.
In the case that $s$ decays dominantly into a bottom quark pair, which is indeed the case for
the parameter space of our interest, we have $R_s^{Zb\bar b} \simeq |C^s_V|^2$, and
thus the LEP excess can be explained if
\bea
0.1 \,\lesssim \,\theta^2_2 \,\lesssim\, 0.25, \quad
m_s\,\simeq\, 98 \, {\rm GeV}.
\eea

Let us examine how $\{\lambda, \mu, m_0\}$ change on the $(\theta_2,\tan\beta)$ plane
for the case that $s$ explains the LEP excess around 98~GeV, and $h$ has the properties
observed at the LHC.
Imposing the condition (\ref{RVV-soln}) for $R^{VV}_h\approx1$, together with $m_s=98$~GeV
and $m^2_H\gg v^2$, one finds that $\lambda$ and $\mu$ can be determined in terms of
$\theta_2$, $\theta_3$, and $\tan\beta$ according to
\bea
\lambda^2 &\approx& 0.27
-0.10\tan\beta\,\theta_2\theta_3
\nonumber \\
&&
+\, 1.11 \left(\frac{m_H}{350\,{\rm GeV}}\right)^2
\Big( 1 -\frac{m^2_h}{m^2_H} \Big) \left(\theta_2^2
-\frac{1.4\tan^2\beta +1.7}{\tan\beta}\theta_2\theta_3\right),
\label{approx-lambda}
\\
\label{approx-mu}
\mu
&\approx& 100\,{\rm GeV}
\left( -0.43 \theta_2 + 16.2 \left(\frac{m_H}{350\,{\rm GeV}}\right)^2
\frac{\theta_3}{\tan\beta} \right)\left(\frac{\lambda}{0.4}\right)^{-1},
\eea
in the expansion in powers of $\theta_i$.
Here we have taken into account that the charged Higgs scalar, whose mass is similar
to $m_H$, should be heavier than about $350$~GeV to satisfy the $b\to s\gamma$ constraint,
barring cancellation with other superparticle contributions \cite{Gambino:2001ew}.
Similarly, one also finds
\bea
\label{approx-m0}
\frac{m^2_0}{m^2_h}
&\approx& 1
-0.4\left( \theta_2^2 - \frac{2}{\tan\beta}\theta_2\theta_3 \right)
\nonumber \\
&&
-\,\frac{11}{\tan^2\beta}
\left(\frac{m_H}{350\,{\rm GeV}}\right)^2
\Big( 1 -\frac{m^2_h}{m^2_H} \Big)
\left(\theta_2^2
-\frac{1.4\tan^2\beta +1.7}{\tan\beta}\theta_2\theta_3\right).
\eea
Although a naive approximation, the above relations help us to qualitatively understand
the effect of scalar mixing for a given value of $m_H$.
For $105\,{\rm GeV}\lesssim m_0 \lesssim 120\,{\rm GeV}$, the last term in (\ref{approx-m0})
cannot be large, constraining Higgs mixing to be
\bea
0 \,\lesssim\,
\delta_\theta\,\equiv\,\theta_2^2 -\frac{1.4\tan^2\beta +1.7}{\tan\beta}\theta_2\theta_3
\,\lesssim\,
0.04 \left(\frac{m_H}{350\,{\rm GeV}}\right)^{-2} \tan^2 \beta.
\label{delta}
\eea
Combining this with the relation (\ref{approx-mu}), the chargino mass bound $|\mu|\gtrsim 100$~GeV
leads to an upper bound on $\tan\beta$,
\bea
\tan\beta &\lesssim& 1.9
\left(\frac{m_H}{350\,{\rm GeV}}\right)
\left(\frac{\theta^2_2}{0.25}\right)^{1/4}
\left(\frac{\lambda}{0.4}\right)^{-1/2}.
\label{mu_tb}
\eea
One can see that the perturbativity bound $\lambda \lesssim 0.7$ can be easily satisfied for
the scalar mixing angles and $\tan\beta$ satisfying (\ref{delta}) and (\ref{mu_tb}).
Note also that, because of the constraint $(\ref{delta})$, \emph{positive} $\theta_2\theta_3$
is favored for the mixing angle $\theta_2$ to be sizable.

\begin{figure}[t]
\begin{center}
\begin{minipage}{16.4cm}
\centerline{
{\hspace*{0cm}\epsfig{figure=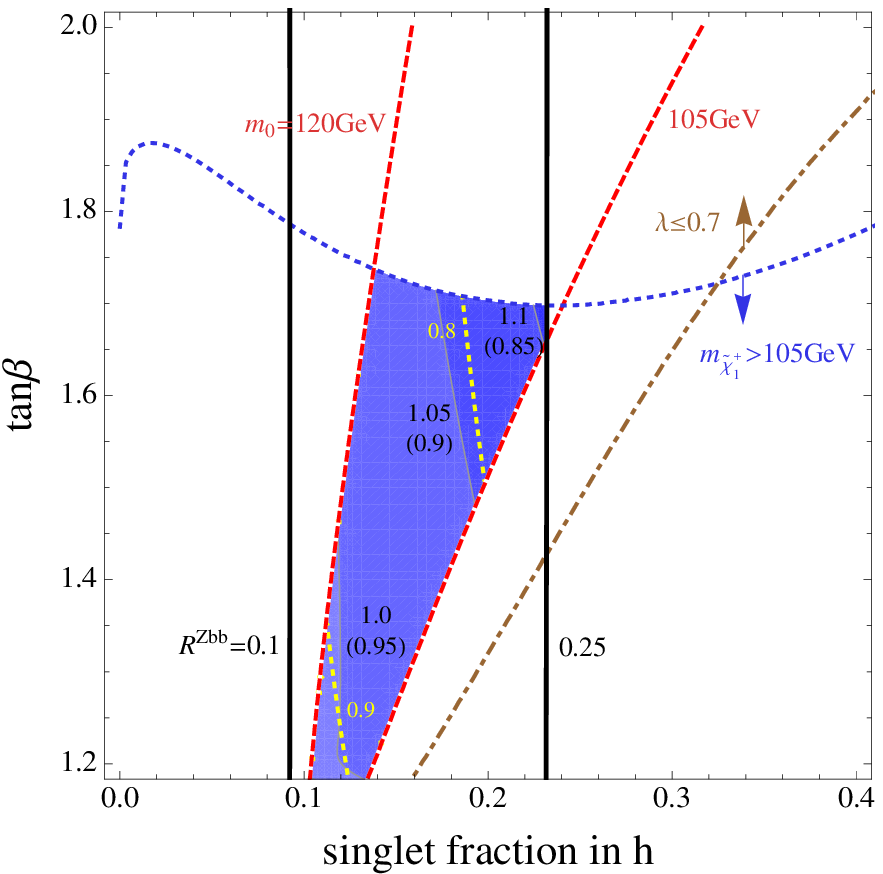,angle=0,width=7.4cm}}
{\hspace*{.4cm}\epsfig{figure=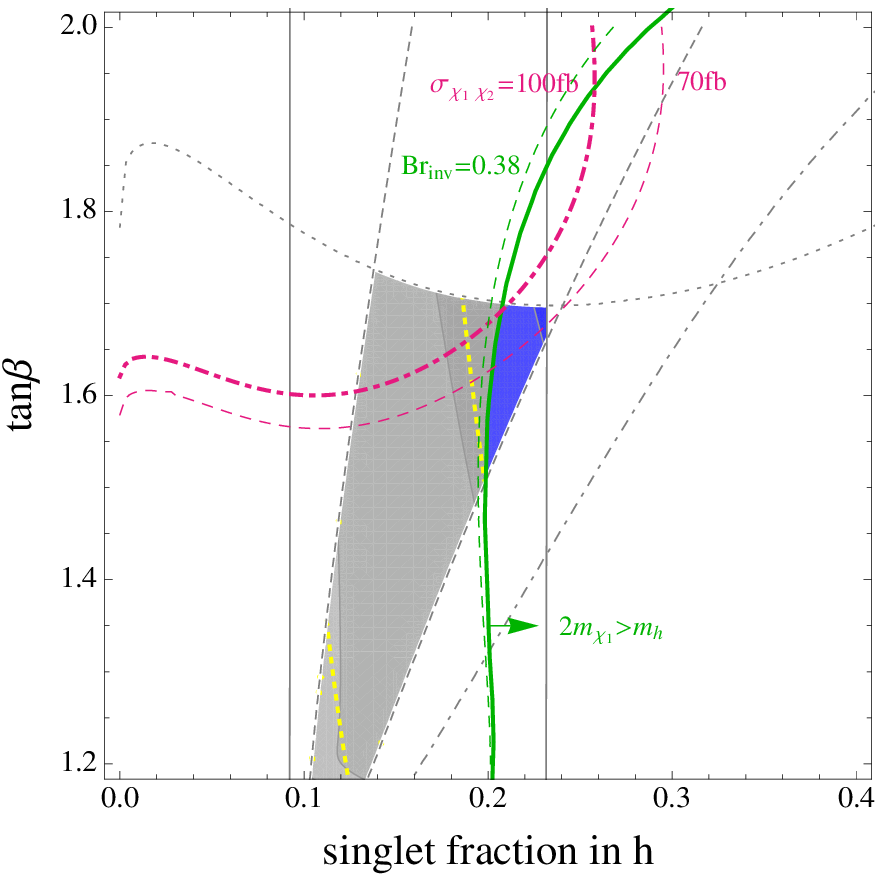,angle=0,width=7.4cm}}
}
\caption{
Singlet fraction ($=c_{\theta_1}^2 s_{\theta_2}^2$) of the SM-like Higgs boson
and $\tan\beta$ consistent with $(a)$ $m_h\simeq 125$~GeV with
$105\,{\rm GeV}\leq m_0 \leq 120\,{\rm GeV}$ (the region between the two dashed red lines),
$(b)$ $R^{VV}_h\approx 1$,
$(c)$ $\lambda\leq 0.7$ (the region above the dot-dashed brown line), and
$(d)$ the LEP bound on the chargino mass (the region below the dotted blue line).
We also impose $0.1\leq R_s^{Zb\bar b}\leq 0.25$ to fit the LEP excess of $Zb\bar b$ at
$m_{b\bar b}\simeq 98$~GeV (the region between the two vertical black lines).
The left panel is the result for the general NMSSM, where we have taken $m_s=98$~GeV,
$m_H=350$~GeV, $|\theta_3|=0.1$ with $\theta_2\theta_3>0$, and $2M_1=M_2=300$~GeV.
The contours of the Higgs signal strengths $R^{\gamma\gamma}_h/R^{VV}_h$
(thin gray line) and $R^{bb}_h/R^{VV}_h$ (dashed yellow line) are depicted, where
the number in the bracket is the di-photon rate for the opposite sign of $\theta_2$.
The right panel shows a viable region of the minimal PQ-NMSSM, where the mixing is
further constrained by $2m_{\chi^0_1}>m_h$ (the right side of the solid green line)
and $\sigma(e^+e^-\to\chi^0_2\chi^0_1)<100\,{\rm fb}$
(the region below the thick-dot-dashed magenta line).
Note that the branching fraction of the Higgs invisible decay, if kinematically open,
is smaller than 0.38 only in the narrow region between the dashed and solid green lines.}
\label{fig:1}
\end{minipage}
\end{center}
\end{figure}

The left plot of Fig.~\ref{fig:1} illustrates the range of the singlet fraction
($=c^2_{\theta_1}s^2_{\theta_2}$) of the SM-like Higgs boson and $\tan\beta$
for which the LEP excess of  $Zb\bar b$ at $m_{b\bar b}\simeq 98$~GeV is explained
by a singlet-like scalar $s$, while satisfying the perturbativity bound on $\lambda$,
the LEP bound on chargino mass, and $105\,{\rm GeV}\leq m_0 \leq 120\,{\rm GeV}$.
Here we have imposed the relation (\ref{RVV-soln}) to have $R^{VV}_h\approx 1$,\footnote{
One may consider a case where the Higgs signal rate into $WW/ZZ$ deviates from the SM value
by an amount $\delta R^{VV}_h$ due to scalar mixing.
Then, the relation (\ref{RVV-soln}) should be modified as
\bea
\theta_1 \approx \frac{\tan\beta}{1.4\tan^2\beta+1.7}
\left(\theta^2_2+ \delta R^{VV}_h \right),
\eea
and the mixing effects can be examined by taking the replacement
$\theta^2_2\to \theta^2_2 + \delta R^{VV}_h$ in the relations (\ref{approx-lambda}) and (\ref{approx-m0}).
As a result, the region consistent with  $105\,{\rm GeV} \leq m_0\leq 120$~GeV will move horizontally
to the left (right) in Fig.~\ref{fig:1} if $\delta R^{VV}_h$ is positive (negative).}
and used $|\theta_3|=0.1$, $m_H=350$~GeV, and the gaugino masses $2M_1=M_2=300$~GeV for the purpose
of illustration.
The LEP bound on the chargino mass puts a lower bound on $|\mu|$, which can be relaxed if the wino
mass $M_2$ is around a few hundred GeV and $\mu M_2<0$.
One can see that the allowed blue-shaded region is determined mainly by the constraints associated
with $\mu$ and $m_0$, whose characteristic features can be understood by the relation (\ref{approx-mu})
and (\ref{approx-m0}).
Note that $\tan\beta$ is bounded from above by the constraint on $|\mu|$ according to (\ref{mu_tb}),
while the constraint on $m_0$ explains the allowed range of $\theta_2^2$ for a given value of $\theta_3$
and $\tan\beta$ through the relation (\ref{delta}).
The allowed region becomes smaller if one increases the wino mass or changes its sign, because then the lower
bound on $\mu$ from the chargino mass bound is strengthened.
We also present in Fig.~\ref{fig:1} the contours of $R^{ii}_h/R^{VV}_h$ for the $b\bar b$ ($\tau\bar\tau$)
and di-photon channels.
For the scalar mixing giving $R^{VV}_h \approx 1$, the $b\bar b$ ($\tau\bar\tau$) signal rate is always below
the $WW/ZZ$ signal rate.
On the other hand, the di-photon signal rate can be either below or above the $WW/ZZ$ signal rate
depending on the sign of the singlet-doublet mixing angle $\theta_2$, because the Higgs
coupling to photons receives a sizable contribution from the charged-higgsino loop when the higgsino mass is around the weak scale.\footnote{
The Higgs coupling to photons receives a loop contribution also from the $h\tilde H^+ \tilde W^-$
interaction, which becomes important when both higgsinos and winos have masses not much above
the weak scale.
Such an effect has been included in our analysis.}

Now we impose the additional constraints that are particularly relevant for the minimal PQ-NMSSM
which predicts a light singlino-like neutralino:
\bea
2m_{\chi^0_1}>m_h, \quad
\sigma(e^+e^-\to \chi^0_2\chi^0_1)\lesssim 100\,{\rm fb}.
\eea
As shown in the right plot of Fig~\ref{fig:1}, only a small region remains viable,
in which $R^{bb}_h/R^{VV}_h$ is about 0.8, and $R^{\gamma\gamma}_h/R^{VV}_h$
deviates from one by about $\pm 0.1$ depending on the sign of $\theta_2$.
On the other hand, notice that the branching fraction for the Higgs invisible
decay is larger than 0.38 in most of the parameter space where the decay mode is kinematically
allowed, except in the narrow region between the dashed and solid green lines where
there is a large kinematic suppression.
A notable feature of the minimal PQ-NMSSM is that a phenomenologically viable parameter
region points toward stops around or below 1 TeV and light higgsinos around the weak scale.

\begin{figure}[t]
\begin{center}
\begin{minipage}{16.4cm}
\centerline{
{\hspace*{0cm}\epsfig{figure=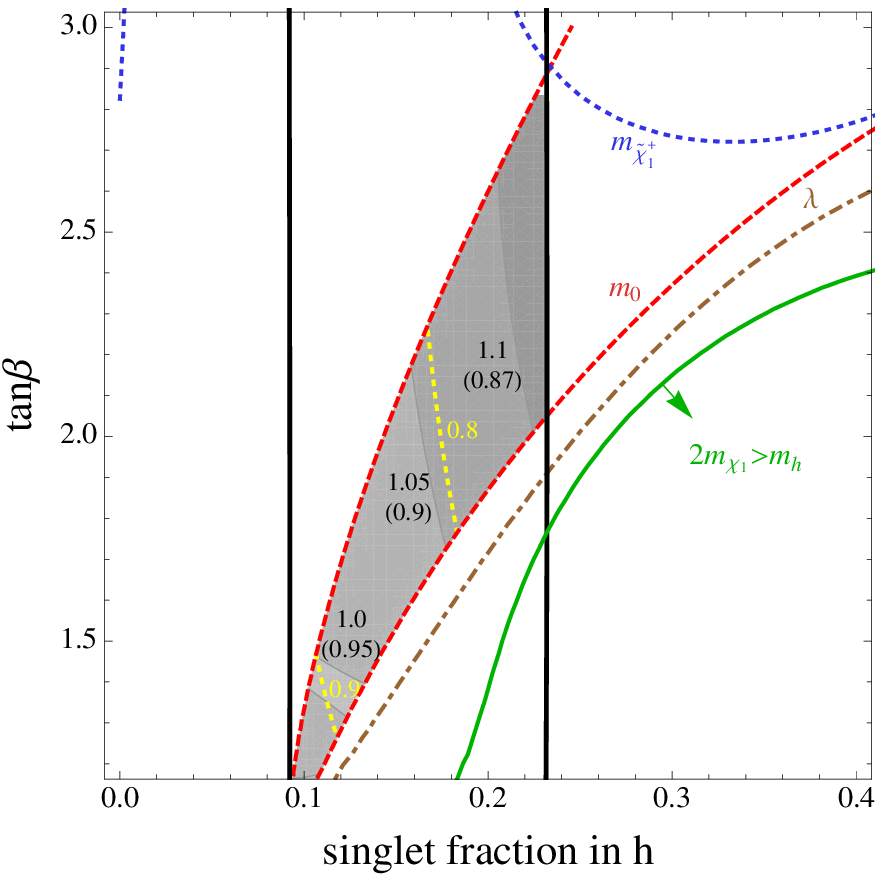,angle=0,width=7.4cm}}
{\hspace*{.4cm}\epsfig{figure=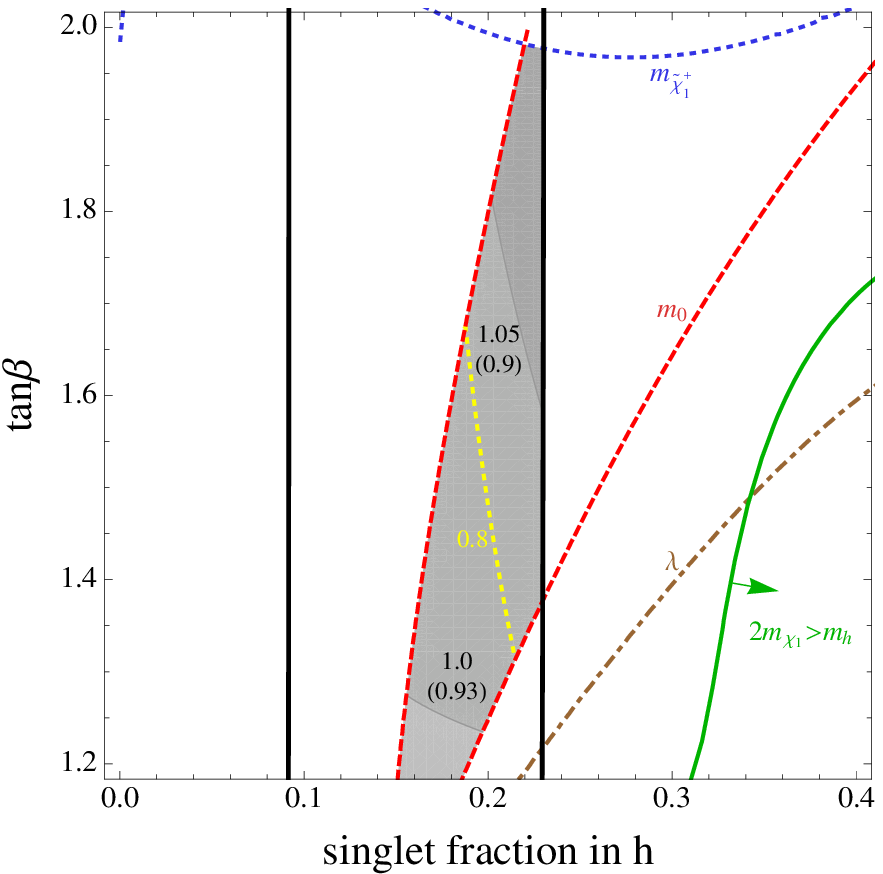,angle=0,width=7.4cm}}
}
\caption{
Higgs mixing consistent with various constraints, where the notations are the same
as in Fig.~\ref{fig:1}.
The left panel shows an allowed region for the general NMSSM with $m_H=500$~GeV
and $|\theta_3|=0.1$, while the right panel is for $m_H=350$~GeV and $|\theta_3|=0.12$.
In both cases, we have taken $m_s=98$~GeV, $\theta_2\theta_3>0$ and $2M_1=M_2=300$~GeV.
For heavier $m_H$ (500~GeV) or larger $|\theta_3|$ (0.12),
there is no viable parameter region satisfying $2m_{\chi_1^0}>m_h$ or
$\rm{Br}(h\rightarrow \chi_1^0 \chi_1^0) < 0.38$ in the minimal PQ-NMSSM.}
\label{fig:2}
\end{minipage}
\end{center}
\end{figure}

Fig.~\ref{fig:2} shows how the allowed region changes with $m_H$ and $\theta_3$.
The left panel is obtained by taking a heavier $m_H$ compared to Fig.~\ref{fig:1},
while the right panel is the result for a larger value of $|\theta_3|$.
The relation (\ref{mu_tb}) indicates that the upper bound on $\tan\beta$ increases as $H$ becomes
heavier, while the relation (\ref{delta}) explains why the shaded region in the figure is reduced for
heavier $m_H$ and why it moves to the right when one takes larger $|\theta_3|$.
On the other hand, for a given $\tan\beta$, $\mu$ becomes large if one raises $m_H$ or $\theta_3$
as can be seen in (\ref{approx-mu}).
A large $\mu$ makes it more difficult to satisfy the condition $2m_{\chi^0_1}>m_h$, so
a phenomenologically viable region of the minimal PQ-NMSSM gets smaller, or disappears,
as $m_H$ or $\theta_3$ increases.

We close this subsection by pointing out that the minimal PQ-NMSSM requires stops around or
below 1~TeV, and higgsinos around the weak scale.
If $m_0$ is larger than about 110~GeV, it is difficult to have $2m_{\chi^0_1}>m_h$, which would
be necessary to forbid $h\rightarrow \chi^0_1\chi^0_1$.
This means that $h$ can be identified as the SM-like Higgs boson observed at the LHC only when
stops are not significantly heavier than 1~TeV.
In addition, combined with $m_0\gtrsim 105$~GeV, the requirement $2m_{\chi^0_1}>m_h$ constrains
$\mu$ to be around the weak scale.
As we will see in the next subsection, these features hold also for the case that $s$ is heavier
than $h$.

\subsection{Singlet-like Higgs boson above 125 GeV}

Let us move to the case where the singlet-like Higgs boson $s$ is heavier than the SM-like Higgs boson $h$.
One of the main differences from the opposite case with $m_s<m_h$ is that the $h$-$s$ mixing always
decreases $m_h$.
Thus we need $\lambda>m_Z/v$ and low $\tan\beta$ in order to arrange $m_h\simeq 125$~GeV
in the presence of scalar mixing, unless $m_0$ is larger than $125$~GeV.
It is clear that the singlet-doublet mixing angle $\theta_2$ can be sizable if $s$ is not much heavier
than $h$.
As in the previous case, the effect of scalar mixing can be understood qualitatively by using
the approximated relations (\ref{approx-lambda}) and (\ref{approx-m0}) after multiplying the second term
with $(m^2_s-m^2_h)/((98{\rm GeV})^2-m^2_h)$, and the relation (\ref{approx-mu})
after multiplying the first term with $m^2_s/(98{\rm GeV})^2$.
Then it follows that a viable region for the general NMSSM with $m_s>m_h$ appears at lower $\tan\beta$
compared to the case with $m_s=98$~GeV.
Hence, it becomes relatively easy to satisfy the condition $2m_{\chi^0_1}>m_h$ in the minimal PQ-NMSSM
where $m_{\chi^0_1}$ is proportional to $\sin2\beta$.

\begin{figure}[t]
\begin{center}
\begin{minipage}{16.4cm}
\centerline{
{\hspace*{0cm}\epsfig{figure=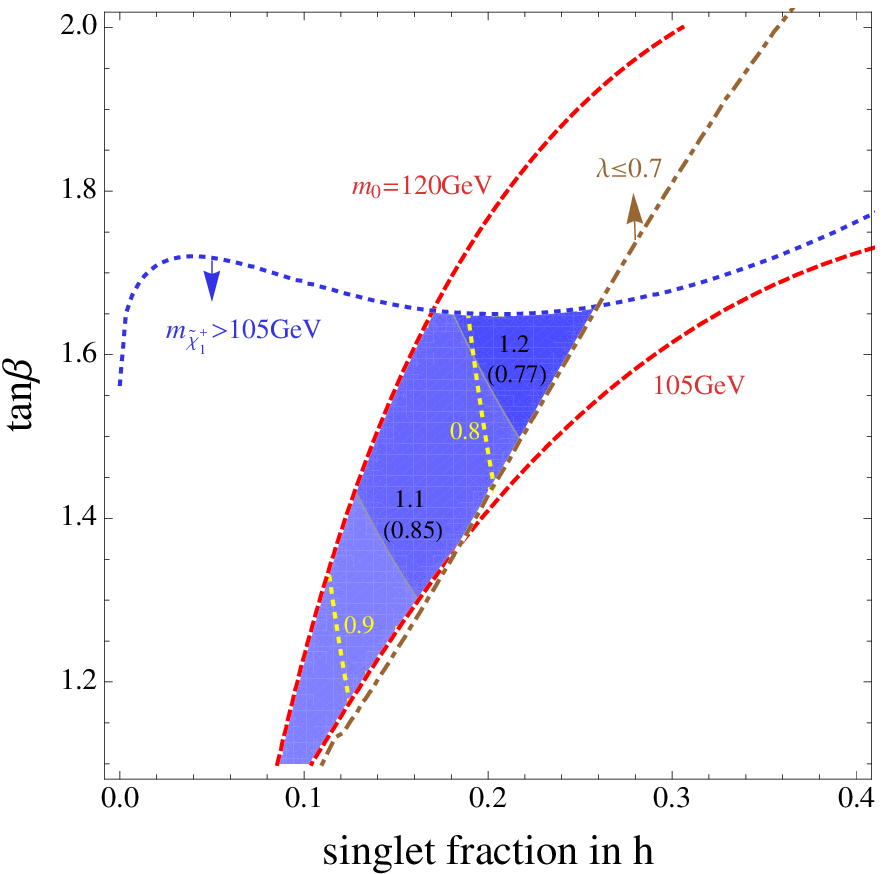,angle=0,width=7.4cm}}
{\hspace*{.4cm}\epsfig{figure=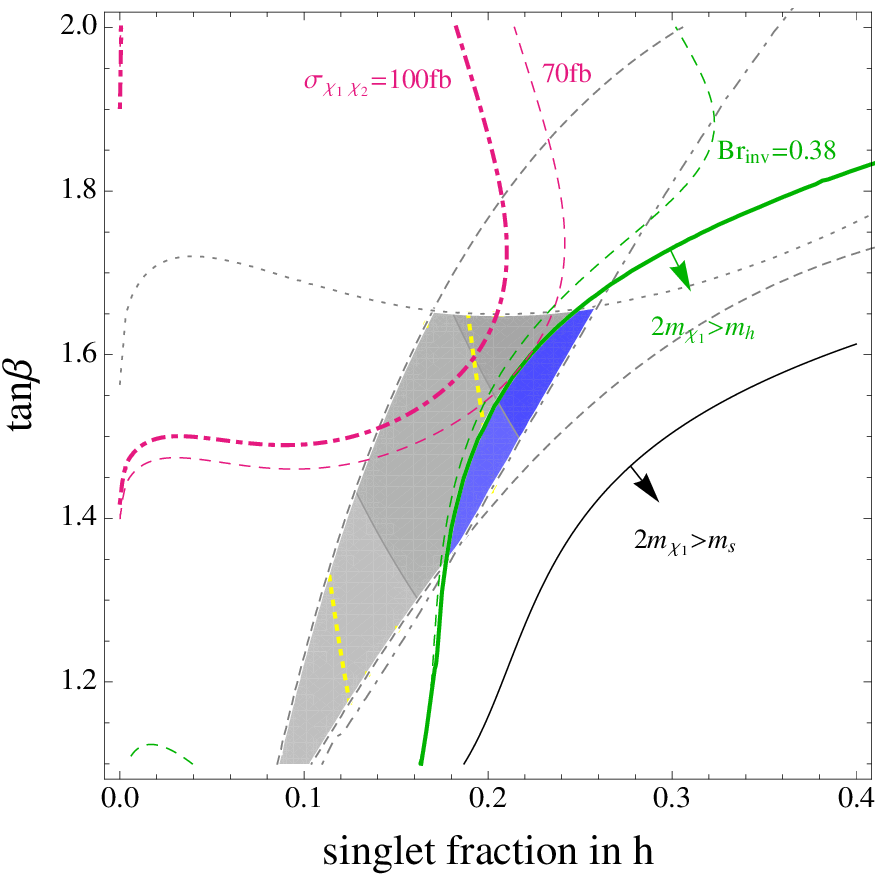,angle=0,width=7.4cm}}
}
\caption{
Higgs mixing consistent with various constraints, where the notations are the same
as in Fig.~\ref{fig:1}.
The left panel shows a viable region in the general NMSSM with $m_s=150$~GeV,
$m_H=350$~GeV, and $|\theta_3|=0.1$ with $\theta_2\theta_3>0$, while the right panel is
the result for the minimal PQ-NMSSM.
Here the gauginos are assumed to be much heavier than the weak scale.
}
\label{fig:3}
\end{minipage}
\end{center}
\end{figure}

In the left panel of Fig.~\ref{fig:3}, we show the region of $(c_{\theta_1}^2s_{\theta_2}^2, \tan\beta)$
compatible with the constraints on $\{\lambda, \mu, m_0\}$ for the general NMSSM with
$m_s=150$~GeV, $m_H=350$~GeV, and $|\theta_3|=0.1$.
Here, for simplicity, we have assumed that the gauginos are much heavier than the weak scale
so they are decoupled well from the singlino-like neutralino.
The figure shows that, for $m_s>m_h$, the perturbativity bound on $\lambda$ is as
important as the other constraints.
However a sizable singlet fraction of $h$ is still allowed.
The right panel shows the result for the minimal PQ-NMSSM, which is further constrained by
the bound on the Higgs invisible decay rate and the LEP bound on the neutralino production
rate.
Again, there is an allowed region with sizable $\theta_2$, in which
$R^{bb}_h/R^{VV}_h$ is around 0.8 for the scalar mixing consistent with $R^{VV}_h\approx 1$,
while $R^{\gamma\gamma}_h/R^{VV}_h$ deviates from one by about $\pm 0.2$ depending
on the sign of $\theta_2$.

\begin{figure}[t]
\begin{center}
\begin{minipage}{16.4cm}
\centerline{
{\hspace*{0cm}\epsfig{figure=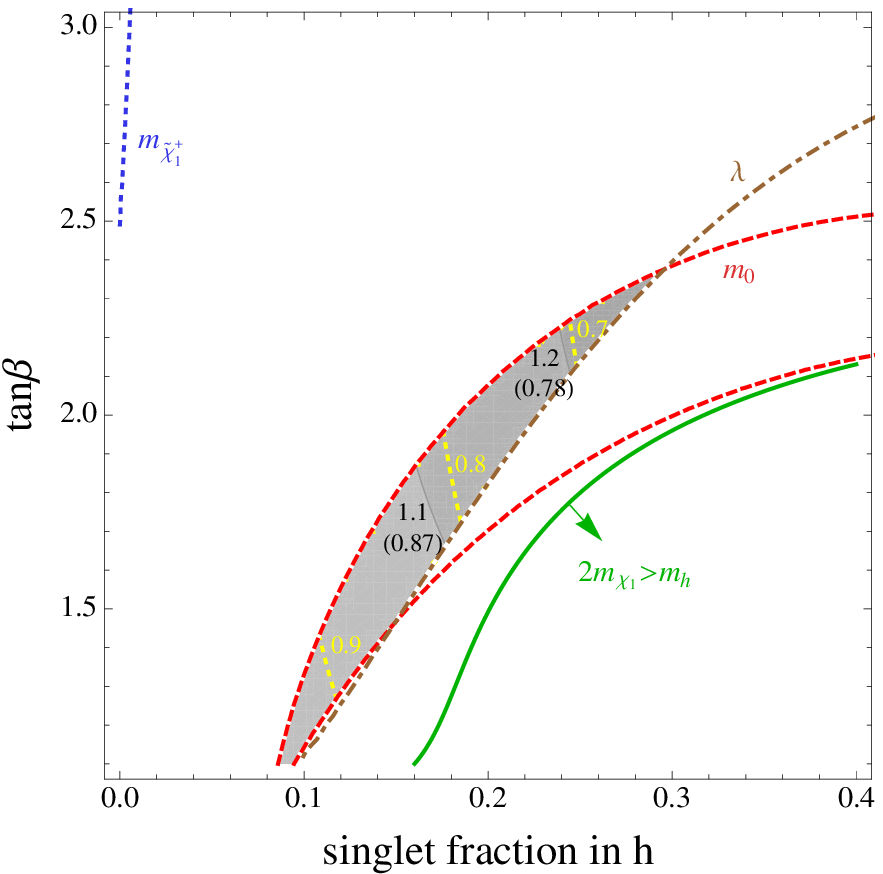,angle=0,width=7.4cm}}
{\hspace*{.4cm}\epsfig{figure=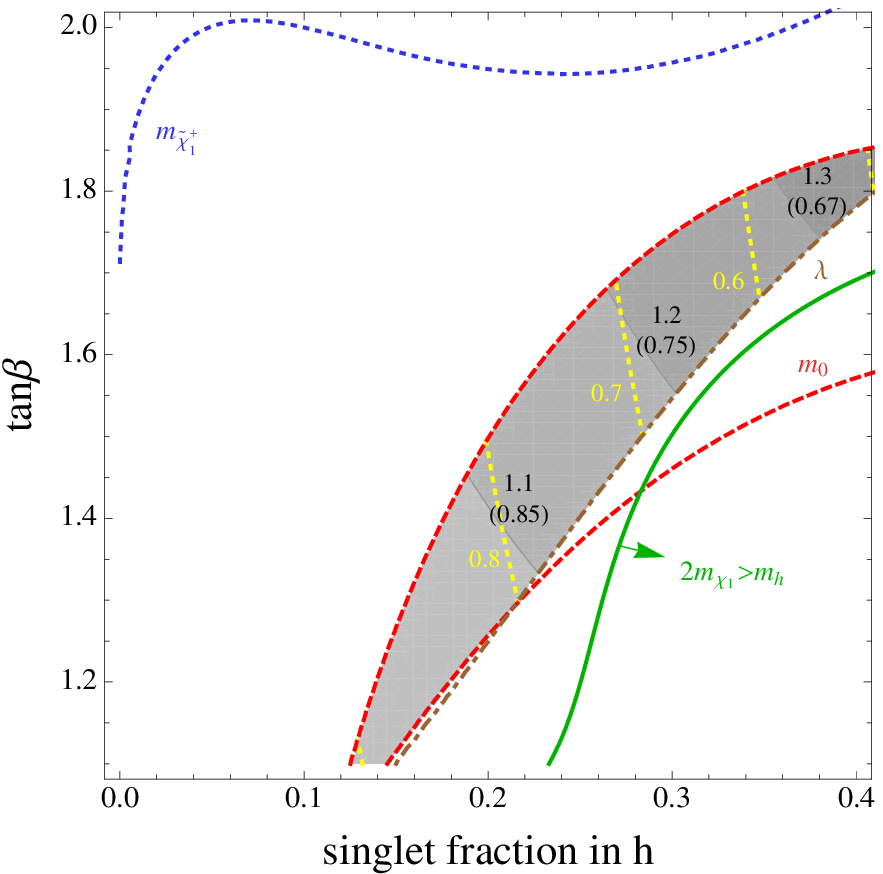,angle=0,width=7.4cm}}
}
\caption{
Higgs mixing consistent with various constraints, where the notations are the same
as in Fig.~\ref{fig:1}.
The left panel shows an allowed region in the general NMSSM with $m_H=500$~GeV and $|\theta_3|=0.1$,
while the right panel is for $m_H=350$~GeV and $|\theta_3|=0.12$.
In both cases, we have taken $m_s=150$~GeV, $\theta_2\theta_3>0$, and assumed the heavy gauginos.
The minimal PQ-NMSSM with heavier $m_H$ (500~GeV) or larger $|\theta_3|$ (0.12)
does not have a viable parameter region satisfying $2m_{\chi_1^0}>m_h$ or
$\rm{Br}(h\rightarrow \chi_1^0 \chi_1^0) < 0.38$.}
\label{fig:4}
\end{minipage}
\end{center}
\end{figure}

On the other hand, if the singlet-like Higgs becomes heavier, only a smaller value of $\theta_2$
will be allowed.
For instance, the singlet fraction of $h$ should be less than about 0.1 for $m_s>250$~GeV
and $m_H=350$~GeV.
Fig.~\ref{fig:4} illustrates how the allowed region of $(c_{\theta_1}^2 s_{\theta_2}^2,\tan\beta)$
varies when $m_H$ becomes heaver (left panel) or $\theta_3$ becomes larger (right panel).
We can see that the allowed region varies in the same way as explained in the previous subsection,
so that the phenomenologically viable region of the minimal PQ-NMSSM becomes smaller
or vanishes as $m_H$ or $\theta_3$ gets larger.

It is worth noting that, if $s$ is heavy enough, it can dominantly decay into a pair of
the lightest neutralino.
For instance, for the minimal PQ-NMSSM depicted in the right panel of Fig.~\ref{fig:3},
the region above the thin black line gives $m_s>2m_{\chi^0_1}$, for which the invisible decay
$s\rightarrow \chi^0_1\chi^0_1$ is open.
This region covers all the parameter space satisfying the phenomenological constraints
discussed here.
We find that the branching fraction of the invisible decay $s\rightarrow \chi^0_1\chi^0_1$ is
about $0.7-0.8$ over the viable region, which would make it difficult to discover $s$ at collider experiments.

\section{Conclusions}\label{sec:conclusions}

It is an interesting possibility that there exist additional light Higgs bosons near the weak scale
other than the SM-like Higgs boson recently discovered at the LHC.
The next-to-minimal supersymmetric standard model (NMSSM) is a natural place to realize this scenario
because it can accommodate a SM-like 125 GeV Higgs boson with less fine-tuning, and additional scalar
bosons can have a light mass without causing a further fine-tuning.
The axion solution to the strong CP problem can be easily incorporated in the NMSSM.
Then an appealing scenario is that an intermediate PQ scale $v_{\rm PQ}\sim \sqrt{m_{\rm soft}M_{Pl}}$
emerges through an interplay between SUSY breaking effect and Planck-scale suppressed effect,
and low energy mass parameters of ${\cal O}(m_{\rm soft})$ in the effective superpotential
are induced by a spontaneous breakdown of the PQ symmetry.

We have examined the Higgs phenomenology in such a PQ-invariant NMSSM while focusing on
the phenomenological consequences of scalar mixing.
The observed mass and signal rates of the SM-like Higgs boson and the LEP bound on the chargino mass
provide important constraints on scalar mixing.
We imposed also the perturbativity bound on the singlet Yukawa coupling $\lambda$, and assumed
stop masses between 600~GeV and a few TeV.
In addition to these constraints, the minimal PQ-NMSSM is further constrained due to the presence
of a light singlino-like neutralino in the model, most notably by the invisible decay
of the SM-like Higgs boson into a neutralino pair.

The singlet-doublet mixing can also affect the mass of the SM-like Higgs boson $h$.
For $m_s < m_h$, we have explored the possibility that the $2\sigma$ excess of the LEP $Zb\bar b$ events
at $m_{b\bar b}\simeq 98$~GeV is explained by a singlet-like Higgs boson $s$
with $m_s \simeq 98$~GeV within the framework of the minimal PQ-NMSSM.
Interestingly enough, this requires a low $\tan\beta$ smaller than about 2, together with stops
around or below 1~TeV, and light doublet-higgsinos around the weak scale.
On the other hand, for the case with $m_s>m_h$, we found that $s$ dominantly decays into
neutralinos in most of the phenomenologically viable parameter region.
In both cases, the signal rate of the SM-like Higgs boson decaying into $b\bar b$ or $\tau\bar\tau$ is
reduced by $10$--$20\%$ compared to the signal rate into $WW/ZZ$, while the signal rate
in the di-photon channel is reduced or enhanced by a similar amount depending on the sign
of the singlet-doublet mixing angle.
Here scalar mixing has been constrained by the observational requirement that
the Higgs signal rate into $WW/ZZ$ be close to the SM value.

\section*{Acknowledgment}
KC is supported by the National Research Foundation of Korea (NRF) grant
(No. 2012R1A2A2A05003214) and the BK21 project funded by the
Korean Government (MEST).
SHI is supported by the NRF of Korea No. 2011-0017051.
KSJ is supported by Grant-in-Aid for Scientific Research (C) (No. 23540283),
and Scientific Research on Innovative Areas (No.23104008).
MS is supported by Basic Science Research Program through the National
Research Foundation of Korea (NRF) funded by the Ministry of Education,
Science and Technology (No. 2011-0011083).

\appendix

\section{Scalar masses and mixings in the NMSSM }\label{appendix}

For the CP-even neutral Higgs bosons defined as
\bea
\hat h &=& \sqrt2\left(
({\rm Re}H^0_d - v\cos\beta)\cos\beta
+ ({\rm Re}H^0_u - v\sin\beta)\sin\beta \right),
\nonumber \\
\hat H &=& \sqrt2\left(
({\rm Re}H^0_d - v\cos\beta)\sin\beta
-({\rm Re}H^0_u - v\sin\beta)\cos\beta \right),
\nonumber \\
\hat s &=& \sqrt2\left(
{\rm Re}S - \langle S \rangle \right),
\eea
the scalar mass matrix is given by
\bea
\hat M^2 =
\left(%
\begin{array}{ccc}
  m^2_{\hat h} & \frac{1}{2}(m^2_Z-\lambda^2 v^2)\sin4\beta &
  \lambda v (2\mu-\Lambda \sin2\beta) \\
  \frac{1}{2}(m^2_Z-\lambda^2 v^2)\sin4\beta & m^2_{\hat H} &
  \lambda v \Lambda \cos2\beta \\
  \lambda v (2\mu-\Lambda \sin2\beta) & \lambda v \Lambda \cos2\beta
  & m^2_{\hat s} \\
\end{array}%
\right),
\label{M2hat}
\eea
where  $\langle |H^0_u| \rangle = v\sin\beta$ and $\langle|H^0_d|\rangle = v\cos\beta$, and
\bea
m^2_{\hat h} &=& m_Z^2\cos^2 2\beta  + \lambda^2 v^2\sin^2 2\beta
+ \frac{3m^4_t}{4\pi^2v^2} \ln\left(\frac{m^2_{\tilde t}}{m^2_t}\right)
+ \frac{3m^4_t}{4\pi^2v^2}\left( X^2_t - \frac{1}{12}X^4_t\right),\\
m_{\hat{H}}^2 &=& \frac{2B\mu}{\sin 2\beta}-(\lambda^2v^2-m_Z^2)\sin^22\beta,
\eea
for the stop mixing parameter  $X_t=(A_t-\mu\cot\beta)/m_{\tilde t}$,
and $\Lambda = A_\lambda +\langle \partial^2_S f \rangle$.
Here $m^2_{\hat s}$ depends on the superpotential $f$ for the singlet $S$, and the soft scalar mass term for it.
The mass matrix is diagonalized by an orthogonal matrix,
\bea
U = \left(%
\begin{array}{ccc}
  c_{\theta_1}c_{\theta_2} & -s_{\theta_1} & -c_{\theta_1}s_{\theta_2} \\
  s_{\theta_1}c_{\theta_2}c_{\theta_3}
  -s_{\theta_2}s_{\theta_3} & c_{\theta_1}c_{\theta_3} &
  -c_{\theta_2}s_{\theta_3} -s_{\theta_1}s_{\theta_2}c_{\theta_3}  \\
  s_{\theta_1}c_{\theta_2}s_{\theta_3} + s_{\theta_2}c_{\theta_3} &
  c_{\theta_1}s_{\theta_3} &
  c_{\theta_2}c_{\theta_3}-s_{\theta_1}s_{\theta_2}s_{\theta_3} \\
\end{array}%
\right),
\label{unitary}
\eea
where $c_\theta=\cos\theta$ and $s_\theta=\sin\theta$ for $\theta_i$ in the range
between $-\pi/2$ and $\pi/2$.
Assuming that the mixing angles are small, the mass eigenstate
\bea
h=c_{\theta_1}c_{\theta_2}\hat h - s_{\theta_1} \hat H -c_{\theta_1}s_{\theta_2} \hat s
\nonumber
\eea
can be identified as the observed SM-like Higgs boson with a mass $m_h\simeq 125$~GeV.

The Lagrangian parameters are related to the Higgs mass eigenvalues and mixing angles as follows:
\begin{eqnarray}
\label{param}
\lambda^2 v^2 &=& m^2_Z
+ \frac{1}{\sin4\beta}
\Big( (m^2_H - m^2_s) s_{\theta_2} s_{2\theta_3}
+ 2(m^2_h - m^2_H c^2_{\theta_3} - m^2_s s^2_{\theta_3})
s_{\theta_1}c_{\theta_2}
\Big) c_{\theta_1},
\nonumber \\
\lambda v \mu &=&
-\frac{1}{4} m^2_h c^2_{\theta_1} s_{2\theta_2}
-\frac{1}{4}(m^2_H-m^2_s) s_{\theta_1} c_{2\theta_2}s_{2\theta_3}
\nonumber \\
&& +\, \frac{1}{4} \Big(
(m^2_H-m^2_s s^2_{\theta_1}) s^2_{\theta_3}
- (m^2_H s^2_{\theta_1} - m^2_s) c^2_{\theta_3}
\Big)s_{2\theta_2}
\nonumber \\
&& -\, \frac{\tan2\beta}{4} \Big(
(m^2_H-m^2_s) c_{\theta_2} s_{2\theta_3}
 - 2(m^2_h-m^2_H c^2_{\theta_3} -m^2_s s^2_{\theta_3})
s_{\theta_1}s_{\theta_2}
\Big)c_{\theta_1}, \nonumber \\
\lambda v \Lambda &=& -\frac{1}{2\cos2\beta}
\Big( (m^2_H - m^2_s) c_{\theta_2} s_{2\theta_3}
- 2(m^2_h - m^2_H c^2_{\theta_3} - m^2_s s^2_{\theta_3})
s_{\theta_1}s_{\theta_2}
\Big)c_{\theta_1},
\eea
and
\bea
m_{\hat{h}}^2 &=&
m_h^2 c_{\theta_1}^2c_{\theta_2}^2
+ m_H^2 (s_{\theta_1} c_{\theta_2} c_{\theta_3} - s_{\theta_2} s_{\theta_3})^2
+ m_s^2 (s_{\theta_2} c_{\theta_3} + s_{\theta_1} c_{\theta_2} s_{\theta_3})^2,
\nonumber \\
m_{\hat{H}}^2 &=& m_h^2 s_{\theta_1}^2
+ m_H^2 c_{\theta_1}^2 c_{\theta_3}^2 + m_s^2 c_{\theta_1}^2 s_{\theta_3}^2,
\nonumber \\
m_{\hat{s}}^2 &=& m_h^2 c_{\theta_1}^2 s_{\theta_2}^2
+ m_H^2 (s_{\theta_1} s_{\theta_2} c_{\theta_3} + c_{\theta_2} s_{\theta_3})^2
+ m_s^2 (c_{\theta_2} c_{\theta_3} - s_{\theta_1} s_{\theta_2} s_{\theta_3})^2.
\end{eqnarray}


\begin{thebibliography}{99}


\bibitem{nilles}
  H.~P.~Nilles,
  ``Supersymmetry, Supergravity and Particle Physics,''
  Phys.\ Rept.\  {\bf 110}, 1 (1984);
  H.~E.~Haber and G.~L.~Kane,
  ``The Search for Supersymmetry: Probing Physics Beyond the Standard Model,''
  Phys.\ Rept.\  {\bf 117} (1985) 75.

\bibitem{PQ-mechanism}
  R.~D.~Peccei and H.~R.~Quinn,
  ``CP Conservation In The Presence Of Instantons,''
  Phys.\ Rev.\ Lett.\  {\bf 38}, 1440 (1977);
  ``Constraints Imposed By CP Conservation In The Presence Of Instantons,''
  Phys.\ Rev.\  D {\bf 16}, 1791 (1977).

\bibitem{PQ-review}
For a recent review, see
J.~E.~Kim and G.~Carosi,
``Axions and the Strong CP Problem,''
Rev.\ Mod.\ Phys.\  {\bf 82}, 557 (2010)  [arXiv:0807.3125 [hep-ph]].

\bibitem{axion-scale}
  H.~Murayama, H.~Suzuki and T.~Yanagida,
  ``Radiative breaking of Peccei-Quinn symmetry at the intermediate mass scale,''
  Phys.\ Lett.\ B {\bf 291}, 418 (1992).

\bibitem{axion-thermal}
 K.~Choi, E.~J.~Chun and J.~E.~Kim,
  ``Cosmological implications of radiatively generated axion scale,''
   Phys.\ Lett.\ B {\bf 403}, 209 (1997)  [hep-ph/9608222].

\bibitem{thermal}
D.~H.~Lyth and E.~D.~Stewart,
  ``Cosmology with a TeV mass GUT Higgs,''
  Phys.\ Rev.\ Lett.\  {\bf 75}, 201 (1995)  [hep-ph/9502417];
  ``Thermal inflation and the moduli problem,''
  Phys.\ Rev.\ D {\bf 53}, 1784 (1996)  [hep-ph/9510204];
S.~Kim, W.~-I.~Park and E.~D.~Stewart,
  ``Thermal inflation, baryogenesis and axions,''
  JHEP {\bf 0901} (2009) 015  [arXiv:0807.3607 [hep-ph]];
  K.~Choi, K.~S.~Jeong, W.~-I.~Park and C.~S.~Shin,
  ``Thermal inflation and baryogenesis in heavy gravitino scenario,''
  JCAP {\bf 0911}, 018 (2009)
  [arXiv:0908.2154 [hep-ph]].



\bibitem{ATLAS-Higgs}
  G.~Aad {\it et al.}  [ATLAS Collaboration],
  ``Observation of a new particle in the search for the Standard
  Model Higgs boson with the ATLAS detector at the LHC,''
  Phys.\ Lett.\ B {\bf 716}, 1 (2012)  [arXiv:1207.7214 [hep-ex]].

\bibitem{CMS-Higgs}
  S.~Chatrchyan {\it et al.}  [CMS Collaboration],
  ``Observation of a new boson at a mass of 125 GeV with the
  CMS experiment at the LHC,''  Phys.\ Lett.\ B {\bf 716}, 30 (2012)
  [arXiv:1207.7235 [hep-ex]].

\bibitem{Review-NMSSM}
For a review, see
  M.~Maniatis,
  ``The Next-to-Minimal Supersymmetric extension of the Standard Model reviewed,''
  Int.\ J.\ Mod.\ Phys.\  {\bf A25}, 3505-3602 (2010).
  [arXiv:0906.0777 [hep-ph]];
  U.~Ellwanger, C.~Hugonie, A.~M.~Teixeira,
  ``The Next-to-Minimal Supersymmetric Standard Model,''
  Phys.\ Rept.\  {\bf 496}, 1-77 (2010).
  [arXiv:0910.1785 [hep-ph]].

\bibitem{fine-tuning}
L.~J.~Hall, D.~Pinner and J.~T.~Ruderman,
  ``A Natural SUSY Higgs Near 126 GeV,''  JHEP {\bf 1204}, 131 (2012)  [arXiv:1112.2703 [hep-ph]];
  Z.~Kang, J.~Li and T.~Li,
  ``On Naturalness of the MSSM and NMSSM,''  JHEP {\bf 1211}, 024 (2012)  [arXiv:1201.5305 [hep-ph]];
  G.~G.~Ross, K.~Schmidt-Hoberg and F.~Staub,
  ``The Generalised NMSSM at One Loop: Fine Tuning and Phenomenology,''  JHEP {\bf 1208}, 074 (2012)  [arXiv:1205.1509 [hep-ph]];
  T.~Gherghetta, B.~von Harling, A.~D.~Medina and M.~A.~Schmidt,
  ``The Scale-Invariant NMSSM and the 126 GeV Higgs Boson,''  JHEP {\bf 1302}, 032 (2013)   [arXiv:1212.5243 [hep-ph]].

\bibitem{diphoton-enhancement-nmssm}
  U.~Ellwanger,
  ``Enhanced di-photon Higgs signal in the Next-to-Minimal Supersymmetric Standard Model,''
  Phys.\ Lett.\ B {\bf 698}, 293 (2011)  [arXiv:1012.1201 [hep-ph]];
  U.~Ellwanger,
  ``A Higgs boson near 125 GeV with enhanced di-photon signal in the NMSSM,''
  JHEP {\bf 1203}, 044 (2012)  [arXiv:1112.3548 [hep-ph]];
 J.~J.~Cao, Z.~X.~Heng, J.~M.~Yang, Y.~M.~Zhang, and J.~Y.~Zhu,
  ``A SM-like Higgs near 125 GeV in low energy SUSY: a comparative study for MSSM and NMSSM,''
  JHEP {\bf 1203}, 086 (2012)  [arXiv:1202.5821 [hep-ph]];
S.~F.~King, M.~Mühlleitner, R.~Nevzorov and K.~Walz,
  ``Natural NMSSM Higgs Bosons,''  Nucl.\ Phys.\ B {\bf 870}, 323 (2013)  [arXiv:1211.5074 [hep-ph]];
   M.~Badziak, M.~Olechowski and S.~Pokorski,
  ``New Regions in the NMSSM with a 125 GeV Higgs,''
  JHEP {\bf 1306}, 043 (2013)  [arXiv:1304.5437 [hep-ph]].

\bibitem{Diphoton_NMSSM}
  K.~Choi, S.~H.~Im, K.~S.~Jeong, and M.~Yamaguchi,
  ``Higgs mixing and diphoton rate enhancement in NMSSM models,''
  JHEP {\bf 1302}, 090 (2013)  [arXiv:1211.0875 [hep-ph]].


\bibitem{nmssm-higgs}
C.~Cheung, S.~D.~McDermott and K.~M.~Zurek,
  ``Inspecting the Higgs for New Weakly Interacting Particles,''
  JHEP {\bf 1304} (2013) 074  [arXiv:1302.0314 [hep-ph]];
R.~Barbieri, D.~Buttazzo, K.~Kannike, F.~Sala and A.~Tesi,
  ``Exploring the Higgs sector of a most natural NMSSM,''
  Phys.\ Rev.\ D {\bf 87} (2013) 115018  [arXiv:1304.3670 [hep-ph]];
  ``One or more Higgs bosons?,''
  arXiv:1307.4937 [hep-ph].

\bibitem{mu-problem}
J.~E.~Kim and H.~P.~Nilles,
  ``The mu Problem and the Strong CP Problem,''
  Phys.\ Lett.\ B {\bf 138} (1984) 150.

\bibitem{Jeong:2011xu}
 K.~Choi, E.~J.~Chun, H.~D.~Kim, W.~I.~Park and C.~S.~Shin,
  ``The $\mu$-problem and axion in gauge mediation,''
  Phys.\ Rev.\ D {\bf 83} (2011) 123503  [arXiv:1102.2900 [hep-ph]];
  K.~S.~Jeong and M.~Yamaguchi,
  ``Axion model in gauge-mediated supersymmetry breaking and a solution to the $\mu/B\mu$ problem,''
  JHEP {\bf 1107}, 124 (2011)  [arXiv:1102.3301 [hep-ph]].




\bibitem{PQ-NMSSM}
  K.~S.~Jeong, Y.~Shoji and M.~Yamaguchi,
  ``Peccei-Quinn invariant extension of the NMSSM,''
  JHEP {\bf 1204}, 022 (2012)  [arXiv:1112.1014 [hep-ph]];
  ``Singlet-Doublet Higgs Mixing and Its Implications on the Higgs mass in the PQ-NMSSM,''
  JHEP {\bf 1209}, 007 (2012)  [arXiv:1205.2486 [hep-ph]].

\bibitem{Bae:2012am}
  K.~J.~Bae, K.~Choi, E.~J.~Chun, S.~H.~Im, C.~B.~Park and C.~S.~Shin,
  ``Peccei-Quinn NMSSM in the light of 125 GeV Higgs,''
   JHEP {\bf 1211}, 118 (2012)  [arXiv:1208.2555 [hep-ph]].

\bibitem{Belanger:2012}
G.~Belanger, U.~Ellwanger, J.~F.~Gunion, Y.~Jiang, S.~Kraml, and J.~H.~Schwarz,
``Higgs Bosons at 98 and 125 GeV at LEP and the LHC,"
JHEP {\bf 1301}, 069 (2013)
[arXiv:1210.1976 [hep-ph]].


\bibitem{Carmi:2012in}
  D.~Carmi, A.~Falkowski, E.~Kuflik, T.~Volansky and J.~Zupan,
  ``Higgs After the Discovery: A Status Report,''
JHEP {\bf 1210}, 196 (2012)
[arXiv:1207.1718 [hep-ph]].

\bibitem{Okada:1990vk}
  Y.~Okada, M.~Yamaguchi and T.~Yanagida,
  ``Upper bound of the lightest Higgs boson mass in the minimal supersymmetric
  standard model,''
  Prog.\ Theor.\ Phys.\  {\bf 85}, 1 (1991);
J.~R.~Ellis, G.~Ridolfi and F.~Zwirner,
  ``Radiative corrections to the masses of supersymmetric Higgs bosons,''  Phys.\ Lett.\ B {\bf 257}, 83 (1991);
H.~E.~Haber and R.~Hempfling,
  ``Can the mass of the lightest Higgs boson of the minimal supersymmetric model be larger than m(Z)?,''  Phys.\ Rev.\ Lett.\  {\bf 66}, 1815 (1991).

\bibitem{arXiv:0712.2903}
  R.~Barbieri, L.~J.~Hall, A.~Y.~Papaioannou, D.~Pappadopulo, and V.~S.~Rychkov,
  ``An alternative NMSSM phenomenology with manifest perturbative unification,''
  JHEP\ {\bf 0803}, 005  (2008)  [arXiv:0712.2903 [hep-ph]].

\bibitem{arXiv:1108.4338}
  K.~Nakayama, N.~Yokozaki and K.~Yonekura,
  ``Relaxing the Higgs mass bound in singlet extensions of the MSSM,''
  JHEP\ {\bf 1111}, 021  (2011)  [arXiv:1108.4338 [hep-ph]].




\bibitem{Miller:2003ay}
D.~J.~Miller, R.~Nevzorov and P.~M.~Zerwas,
  ``The Higgs sector of the next-to-minimal supersymmetric standard model,''
  Nucl.\ Phys.\ B {\bf 681}, 3 (2004)  [hep-ph/0304049].


\bibitem{PDG}
  K.~Nakamura {\it et al.}  [Particle Data Group],
  ``Review of particle physics,''
  J.\ Phys.\ G {\bf 37}, 075021 (2010).


\bibitem{ATLAS-stop}
ATLAS Collaboration, ATLAS-CONF-2013-024;
ATLAS-CONF-2013-037.

\bibitem{CMS-stop}
CMS Collaboration, CMS-PAS-SUS-12-023.


\bibitem{Arvanitaki:2012}
  A.~Arvanitaki and G.~Villadoro
  ``A Non Standard Model Higgs at the LHC as a Sign of Naturalness,''
JHEP {\bf 1202}, 144 (2012)
[arXiv:1112.4835 [hep-ph]].

\bibitem{Blum:2013}
 K.~Blum, R.~T.~D'Agnolo, and J.~Fan,
  ``Natural SUSY Predicts: Higgs Couplings,''
JHEP {\bf 1301}, 057 (2013)
[arXiv:1206.5303 [hep-ph]].

\bibitem{nMSSM1}
  C.~Panagiotakopoulos and K.~Tamvakis,
  ``New minimal extension of MSSM,''
  Phys.\ Lett.\ B\ {\bf 469}, 145  (1999)  [hep-ph/9908351].

\bibitem{nMSSM2}
  C.~Panagiotakopoulos, A.~Pilaftsis,
  ``Higgs scalars in the minimal nonminimal supersymmetric standard model,''
  Phys.\ Rev.\  {\bf D63}, 055003 (2001).
  [hep-ph/0008268].

\bibitem{nMSSM3}
  A.~Dedes, C.~Hugonie, S.~Moretti and K.~Tamvakis,
  ``Phenomenology of a new minimal supersymmetric extension of the standard
  model,''
  Phys.\ Rev.\  D {\bf 63}, 055009 (2001)
  [arXiv:hep-ph/0009125].

\bibitem{DM-nMSSM}
  A.~Menon, D.~E.~Morrissey and C.~E.~M.~Wagner,
  ``Electroweak baryogenesis and dark matter in the nMSSM,''
  Phys.\ Rev.\  D {\bf 70} (2004) 035005
  [arXiv:hep-ph/0404184].


\bibitem{PH-nMSSM}
  C.~Balazs, M.~S.~Carena, A.~Freitas and C.~E.~M.~Wagner,
  ``Phenomenology of the nMSSM from colliders to cosmology,''
  JHEP\ {\bf 0706}, 066  (2007)  [arXiv:0705.0431 [hep-ph]];
  J.~Cao, H.~E.~Logan and J.~M.~Yang,
  ``Experimental constraints on nMSSM and implications on
  its phenomenology,''  Phys.\ Rev.\ D\ {\bf 79}, 091701  (2009)
  [arXiv:0901.1437 [hep-ph]].

\bibitem{OPAL}
  OPAL collaboration, G. Abbiendi, et al,
  ``Search for Chargino and Neutralino Production at $ \sqrt{s} = 192-209 $ GeV at LEP,"
  Eur. Phys. J. C {\bf 35}, 1 (2004) [hep-ex/0401026].


\bibitem{Giardino:2013bma}
  P.~P.~Giardino, K.~Kannike, I.~Masina, M.~Raidal and A.~Strumia,
  ``The universal Higgs fit,''  arXiv:1303.3570 [hep-ph].

 \bibitem{Belanger:2013}
 G.~Belanger, B.~Dumont, U.~Ellwanger, J.~F.~Gunion, and S.~Kraml,
 ``Global fit to Higgs signal strengths and couplings and implications for extended Higgs sectors,"
 arXiv:1306.2941 [hep-ph].


\bibitem{LEP98GeV}
  S.~Schael {\it et al.}
  [ALEPH and DELPHI and L3 and OPAL and LEP Working Group for Higgs Boson Searches Collaborations],
  ``Search for neutral MSSM Higgs bosons at LEP,''  Eur.\ Phys.\ J.\ C {\bf 47}, 547 (2006)  [hep-ex/0602042];
  R.~Barate {\it et al.}
  [LEP Working Group for Higgs boson searches and ALEPH and DELPHI and L3 and OPAL Collaborations],
  ``Search for the standard model Higgs boson at LEP,''  Phys.\ Lett.\ B {\bf 565}, 61 (2003)  [hep-ex/0306033].

\bibitem{Gambino:2001ew}
  P.~Gambino and M.~Misiak,
  ``Quark mass effects in $\bar B \to X_s\gamma$,''
  Nucl.\ Phys.\ B {\bf 611}, 338 (2001)  [hep-ph/0104034].



\end{thebibliography}
\end{document}